\documentclass[11pt, a4paper]{article}
\usepackage[body={15.4cm,23.6cm}, top=2.3cm]{geometry}

\usepackage{amsmath, amsfonts,amsthm,amssymb,bm}
\usepackage{dsfont}

\usepackage{geometry} 
\usepackage{extarrows}
\usepackage[english]{babel}
\usepackage{setspace}
\usepackage[normalem]{ulem}

\usepackage{latexsym,amssymb,amsfonts,graphicx}
\usepackage{amsmath,dsfont}
\usepackage{verbatim}
\usepackage{mathrsfs}
\usepackage{color}
\usepackage{epsfig}

\usepackage{url}

\usepackage[colorlinks,linkcolor=blue,citecolor=blue,anchorcolor=blue]{hyperref}

\newcommand{\Px}{ \mathbb{P} }

\newcommand{\Ex}{ \mathbb{E} }

\newcommand{\Zx}{\mathbb{Z}}

\def\esssup_#1{\underset{#1}{\Xi}}
\def\essinf_#1{\underset{#1}{\mathrm{ess\,inf\, }}}
\def\argmax_#1{\underset{#1}{\mathrm{arg\,max\, }}}
\def\argmin_#1{\underset{#1}{\mathrm{arg\,min\, }}}

\newcommand{\Gx}{\mathbb{G}}
\newcommand{\Fx}{\mathbb{F} }

\newcommand{\G}{\mathcal{G}}
\newcommand{\R}{\mathds{R}}

\newtheorem{theorem}{Theorem}[section]
\newtheorem{definition}{Definition}[section]

\newtheorem{lemma}[theorem]{Lemma}

\definecolor{Red}{rgb}{1.00, 0.00, 0.00}

\definecolor{DRed}{rgb}{0.5, 0.00, 0.00}

\definecolor{Blue}{rgb}{0.00, 0.00, 1.00}

\definecolor{Green}{rgb}{0.0, 0.4, 0.0}

\definecolor{Magenta}{rgb}{1.0, 0, 1.0}

\allowdisplaybreaks

\title
{
Optimal Credit Investment and Risk Control for an Insurer with Regime-Switching
}

\author{Lijun Bo\thanks{Email: lijunbo@ustc.edu.cn, School of Mathematical Sciences, University of Science and Technology of China, Hefei, Anhui
Province, 230026, China,  and Wu Wen Tsun Key
Laboratory of Mathematics, Chinese Academy of Science, Hefei, Anhui Province, 230026, China.}\and Huafu Liao\thanks{Email: lhflhf@mail.ustc.edu.cn, School of Mathematical Sciences, University of Science and Technology of China, Hefei, Anhui
Province, 230026, China.}\and Yongjin Wang\thanks{Email: yjwang@nankai.edu.cn, Business School, Nankai University, Tianjin, 300071, China.}
}

\begin{document}
\maketitle
\begin{abstract}

This paper studies an optimal investment and risk control problem for an insurer with default contagion and regime-switching. The insurer in our model allocates his/her wealth across multi-name defaultable stocks and a riskless bond under regime-switching risk. Default events have an impact on the distress state of the surviving stocks in the portfolio. The aim of the insurer is to maximize the expected utility of the terminal wealth by selecting optimal investment and risk control strategies. We characterize the optimal trading strategy of defaultable stocks and risk control for the insurer. By developing a truncation technique, we analyze the existence and uniqueness of global (classical) solutions to the recursive HJB system. We prove the verification theorem based on the (classical) solutions of the recursive HJB system.

\vspace{0.3 cm}

{\noindent{\textbf{AMS 2000 subject classifications}: 3E20, 60J20.}

\vspace{0.3 cm}

}
\end{abstract}

\section{Introduction}\label{sec:intro}
Since the seminal works of Merton~\cite{Merton69, Merton71}, portfolio optimization problems have been the subject of considerable investigations.
In recent years, the hybrid diffusion models have received a considerable amount of attention from both researchers and practitioners.
{In particular, the regime-switching model (as a class of hybrid models) is usually proposed to capture the influence on the behavior of the
market caused by transitions in the macroeconomic system or the macroscopic readjustment and regulation. Zhang and Zhou~\cite{zhangzhou09} study
the valuation of stock loan in which the underlying stock price is modeled as a  Markov modulated geometric Brownian motion using a two-state hidden Markov chain.}
Elliott, et al.~\cite{ElliSiu07} consider the pricing of options under a generalized Markov modulated jump diffusion model. Capponi, et al.~\cite{Ago14mf1} obtain
a Poisson series representation for the arbitrage-free price process of vulnerable contingent claims in a market driven by an underlying continuous-time
Markov chain. {Apart from the classical Merton's model of utility maximization on terminal wealth, there has been an increasing consideration of
different stochastic control criteria for portfolio management in recent years.} Zhou and Yin~\cite{zhangyin03} study the Markowitz's mean-variance
portfolio selection with regime-switching in a continuous time model. {Elliott and Siu~\cite{ElliSiu} investigate an optimal portfolio  selection  problem in
a Markov modulated Black-Scholes market when an economic agent faces model uncertainty.}
Shen~and Siu~\cite{ShenSiu} discuss a consumption-portfolio optimization problem in a hidden Markov modulated asset price model
with multiple risky assets under the situation that an economic agent only has access to information about the price processes
of risky shares. Andruszkiewicz, et al.~\cite{Davis16} consider a risk-sensitive investment problem under a jump diffusion regime-switching market model.

{The objective of this paper is to consider an analytical framework for the portfolio allocation and risk control of an insurer, which explicitly accounts for the interaction between regime-switching and credit risk.} These two sources of risk have been identified as tightly linked in empirical research, see, for example, Campbell and Taksler~\cite{Campbell}. For pricing, models accounting for the dependence of default intensities on asset volatilities have been proposed by Carr and Linetsky~\cite{CarrLinetsky}, Carr and Wu~\cite{CarrWu}, and extended to a multi-name context by Mendoza-Arriaga and Linetsky~\cite{MendozaMF}. We propose a model in which switching regimes, capturing the state or modes of the underlying credit market, drive both volatility and default risk of the risky asset price processes. Moreover, the total risk controlled by liabilities of the insurer is driven by the switching regimes and the credit states of the portfolio. Zou and Cadenillas~\cite{ZouCa14} consider an optimal investment and risk control problem with a single default-free asset. The case with multiple default-free assets and regime-switching is extended by Zou and Cadenillas~\cite{ZouCa17}. More recently, Peng and Wang \cite{Pengwang} study the optimal investment strategy and risk control for an insurer who has some inside information on the insurance business. Bo and Wang~\cite{Bowang} focus on an optimal investment and risk control problem for an insurer under stochastic diffusive factors.

We incorporate the interaction between regime-switching and default contagion risk into the risk control model. Differently from the default-free case, default events have an impact on the distress state of the surviving stocks in the portfolio. {Since defaults can occur sequentially, the default intensities of the surviving names are affected by the default events of other stocks in the portfolio. Hence, the HJB system associated with the stochastic control problem is recursive in terms of default states of the portfolio.} The depth of the recursion equals the number of stocks in the portfolio. We analyze the HJB equation and the constrained equation satisfied by the optimal strategy of stocks using a backward recursion. The recursive procedure starts from the state in which all stocks are defaulted and regresses toward the state in which all stocks are alive. Since the policy space of our control problem is not assumed to be compact, the main difficulty in the analysis of solutions to this coupled system lies in the general default state and the non-Lipschitz nonlinearities of the system. Andruszkiewicz, et al.~\cite{Davis16} deal with a risk-sensitive investment problem in a finite-factor model under a compact policy space. The existence and uniqueness of solutions to their HJB equation can be established by verifying the globally Lipschitz-continuous coefficients. We prove in this paper that the nonlinearities of the coupled system are Lipschitz-continuous only when the variable corresponding to the solution is not close to zero (see Lemma~\ref{lem:Gkesti}). This suggests developing a truncation technique such that the truncated nonlinearity in the system is globally Lipschitz-continuous and considering an approximation of the truncated recursive coupled systems. For this purpose, we establish a key comparison result (see Lemma~\ref{lem:comparison}) for two coupled monotone dynamical systems. We refer the reader to Smith~\cite{smith08} for the definition of monotone dynamical systems. In order to construct the limit of the approximating truncated systems, we prove that the approximating systems admit a uniform (strictly positive) lower bound, and then this limit can be verified to be the unique global solution of our recursive HJB system (see Theorem \ref{thm:solutionk}).

The rest of the paper is organized as follows. Section \ref{sec:model} introduces the market model with regime-switching and credit risk interaction. Section \ref{sec:framework} formulates the dynamic optimization problem for an insurer and derives the recursive HJB system. Section~\ref{sec:HJB-sys} analyzes the (classical) solutions of the recursive HJB system. The optimal investment and risk control strategies are characterized in the same section. A verification theorem is also proved in the same section. Section~\ref{sec:numerics} develops a numerical analysis. Additional technical proofs are provided in the Appendix.

\section{The Model}\label{sec:model}

We consider a financial market consisting of $n\geq1$ defaultable stocks and a risk-free money market account.
Let $(\Omega,{\mathcal G},{\Gx},\Px)$ be a complete filtered probability space, where the global filtration
$\Gx:=\Fx \vee {\Zx}_1\vee{\Zx}_2$ is augmented by all $\Px$-null sets so as to satisfy the usual conditions.
Let $T>0$ be the finite target horizon. The filtration $\Fx:=({\mathcal{F}}_t)_{t\in[0,T]}$, where $\mathcal{F}_t$ is the sigma-algebra generated by independent multi-dimensional standard Brownian motions denoted by $W:=(W_j(t);\ j=1,\ldots,d)_{t\in[0,T]}^{\top}$, $\bar{W}:=(\bar{W}_j(t);\ j=1,\ldots,\bar{d})_{t\in[0,T]}^{\top}$ and a regime-switching process $Y:=(Y(t))_{t\in[0,T]}$ introduced below. Here $d,\bar{d}\geq1$ and we use $\top$ to denote the transpose operator.
We next specify the filtrations $\Zx_1$ and $\Zx_2$. The default state is described by an $n$-dimensional default indicator process $Z:=(Z_j(t);\ j=1,\ldots,n)_{t\in[0,T]}$ which takes values on ${\cal S}:=\{0,1\}^n$. For $j=1,\ldots,n$, the default time of the $j$-th stock is given by
\begin{eqnarray}\label{eq:default-time}
\tau_j := \inf\{t\geq0;\ Z_j(t)=1\}.
\end{eqnarray}
The filtration $\Zx_1:=({\mathcal{Z}}_{1t})_{t\in[0,T]}$, where the sigma-algebra ${\cal Z}_{1t}:=\bigvee_{j=1}^n{\sigma(Z_j(s);\ s\leq t)}$.
{Hence $\Zx_1$ contains all information about default events until the target horizon $T$. The filtration $\Zx_2:=({\cal Z}_{2t})_{t\in[0,T]}$ where the sigma-algebra ${\cal Z}_{2t}:=\sigma((N_{i,z}(s),\ (i,z)\in\{1,\ldots,m\}\times{\cal S});\ s\leq t)$. Here $N_{i,z}:=(N_{i,z}(t))_{t\in[0,T]}$ for $(i,z)\in\{1,\ldots,m\}\times{\cal S}$ are independent Poisson processes with respective intensities $\nu(i,z)>0$, which will be used to model the risk control process of an insurer introduced in \eqref{eq:Ct} and \eqref{eq:N} below.} Our model consists of four blocks: the regime-switching process, the credit model, the price processes and the risk process for an insurer. Each of these blocks will be detailed in the sequel.

\noindent{\it Regime-switching process.}\quad The regime-switching process $Y$ here is described as a continuous-time (conservative) Markov chain with state space $\{1,\ldots,m\}$ where $m\geq1$, which is independent of the multi-dimensional Brownian motions $(W,\bar{W})$. {The generator of the Markov chain $Y$ is given by an $m\times m$-dimensional matrix $Q:=(q_{ij})_{m\times m}$. This yields that $q_{ii}\leq0$ for $i\in\{1,\ldots,m\}$, $q_{ij}\geq0$ for $i\neq j$, and $\sum_{j=1}^{m}q_{ij}=0$ for $i\in\{1,\ldots,m\}$ (i.e., $\sum_{j\neq i}q_{ij}=-q_{ii}$ for $i\in\{1,\ldots,m\}$).}

\noindent{\it Credit risk model.}\quad The joint process $(Y,Z)$ of the regime-switching process and the default indicator process is a joint Markov process with state space $\{1,\ldots,m\}\times\mathcal{S}$. Moreover, at any time $t\in[0,T]$, the default indicator process transits from a state $Z(t) :=(Z_1(t),\ldots,Z_{j-1}(t),Z_j(t),Z_{j+1}(t),\ldots,Z_n(t))$ in which the stock $j$ is alive ($Z_j(t)=0$)
to the neighbour state ${Z}^j(t):=(Z_1(t),\ldots,Z_{j-1}(t),1-Z_j(t),Z_{j+1}(t),\ldots,Z_n(t))$ in which the stock $j$ has defaulted at a stochastic rate $\mathds{1}_{Z_j(t)=0}h_{j}(Y(t),Z(t))$. Here $h_j(i,z)>0$ for all $(i,z)\in\{1,\ldots,m\}\times{\cal S}$. We assume that $Y(t)$, $Z_1(t),\ldots,Z_n(t)$ will not jump simultaneously almost surely. Consequently, the default intensity of the $j$-th stock may change either if any other stock in the portfolio defaults (contagion effect), or if there are regime-switchings (market risk effect). Our default model thus belongs to the rich class of interacting intensity models, introduced by Frey and Backhaus \cite{FreyBackhaus04} (see also the interacting default intensity model with diffusive factors introduced in Birge, et al.~\cite{BirgeBoCapponi}). Hereafter, we set $h(i,z):=(h_j(i,z);\ j=1,\ldots,n)^{\top}$ for $(i,z)\in \{1,\ldots,m\}\times{\cal S}$.

\noindent{\it Price processes.}\quad The vector of the price processes of the $n$ defaultable stocks is denoted by $\tilde{S}:=(\tilde{S}_j(t);\ j=1,\ldots,n)_{t\in[0,T]}^{\top}$. For $t\in[0,T]$, the price process of the $j$-th defaultable stock is given by
\begin{equation}
\tilde{S}_j(t)=(1-Z_j(t))S_j(t), \qquad \; j = 1,\ldots,n.
\label{eq:pricedef}
\end{equation}
In other words, the price of the $j$-th stock is given by the predefault price $S_j(t)$ up to ${\tau_j}-$, and  jumps to $0$ at time ${\tau_j}$, where it remains forever afterwards. The dynamics of the pre-default price process $S:=(S_j(t);\ j=1,\ldots,n)_{t\in[0,T]}^{\top}$ of the $n$ defaultable stocks is given by
\begin{align}\label{eq:P}
dS(t) = {diag}(S(t)) [(\mu(Y(t))+h(Y(t),Z(t))) dt + \sigma(Y(t))dW(t)].
\end{align}
Above, ${diag}(S(t))$ is the diagonal $n\times n$-dimensional matrix with diagonal elements $S_j(t)$ for $j=1,\ldots,n$. For each $i\in\{1,\ldots,m\}$, the vector $\mu(i)$ is $\R^n$-valued, and $\sigma(i)$ is an $\R^{n\times d}$-valued matrix such that $\sigma(i)\sigma(i)^{\top}$ is positive definite. Eq.~\eqref{eq:P} indicates that the investor holding the credit sensitive security is compensated for the incurred default risk at the premium rate $h(Y(t),Z(t))$. Using equations~\eqref{eq:pricedef}, \eqref{eq:P} and integration by parts, the dynamics of the defaultable stock prices can be given by
\begin{align}\label{eq:tildeP}
d\tilde{S}(t) = {diag}(\tilde{S}(t)) [\mu(Y(t))dt +  \sigma(Y(t))dW(t)-dM(t)],
\end{align}
where $M:=(M_j(t);\ j=1,\ldots,n)_{t\in[0,T]}^{\top}$ is a pure jump $\Px$-martingale given by
\begin{align}\label{eq:taui}
M_j(t)&:= Z_j(t) - \int_0^{t\wedge\tau_j}h_j(Y(s),Z(s))ds,\ \ \ \ \ \ t\in[0,T].
\end{align}

\noindent{\it Risk control process.}\quad For the risk control process, denote by $\eta(t)$ the $\Gx$-predictable total outstanding number of
policies (liabilities) at time $t$. The risk model for claims is described as an extensive Cram\'er-Lundberg model, in which the claim (risk)
per policy $C=(C(t))_{t\in[0,T]}$ is given by the following dynamics
\begin{align}\label{eq:Ct}
dC(t)=c(Y(t))dt+\phi(Y(t))dW(t)+\bar{\phi}(Y(t))d\bar{W}(t)+g(Y(t-))dN(t),
\end{align}
where, for each $i=1,\ldots,m$, the volatilities $\phi(i)$ and $\bar{\phi}(i)$ are respectively $d$-dimensional and $\bar{d}$-dimensional nonzero row vectors, the drift $c(i)\in\R$, and the positive jump size (claim size) $g(i)\in\R_+:=(0,\infty)$. Here, the jump process $N:=(N(t))_{t\in[0,T]}$ is a Markov modulated Poisson process with positive intensity process given by $(\nu(Y(t),Z(t)))_{t\in[0,T]}$. For $t\in[0,T]$, the process $N(t)$ represents the number of claims occurring in time interval $[0,t]$. More precisely, we can rewrite $N(t)$ as
\begin{align}\label{eq:N}
N(t)=\sum_{(i,z)\in\{1,\ldots,m\}\times{\cal S}}\int_0^t\mathds{1}_{Y(s-)=i,Z(s-)=z}dN_{i,z}(s).
\end{align}
We recall that for $(i,z)\in\{1,\ldots,m\}\times{\cal S}$, $N_{i,z}=(N_{i,z}(t))_{t\in[0,T]}$ are independent Poisson processes with respective intensities $\nu(i,z)$, and moreover they are also independent of the random processes $(W,\bar{W},Y)$. Then, we have that, for $t\in[0,T]$,
\begin{align}\label{eq:tildeN}
\tilde{N}(t) &:=N(t) - \sum_{(i,z)\in\{1,\ldots,m\}\times{\cal S}}\int_0^t\mathds{1}_{Y(s)=i,Z(s)=z}\nu(i,z)ds=N(t)-\int_0^t\nu(Y(s),Z(s))ds
\end{align}
is a $\Px$-martingale. An example of insurance product whose arrival intensity of claims depends on the default states of stocks and the regimes of the economy is so-called Trade Credit Insurance (see, e.g., Jones~\cite{PMJones}). Trade Credit Insurance protects a supplier from the risk of buyer's non-payment. The supplier delivers unpaid goods or services to the buyer and allows a deferred payment from the buyer.  To ensure the payment, the supplier purchases trade credit insurance products. In exchange for the premia, the insurer covers the payment if the buyer defaults. This implies that claims arrive when the buyer fails to pay the suppliers due to credit risk such as protracted default, insolvency, and bankruptcy, etc. Consequently, the probability of buyer's default is correlated with the default states of stocks and the regimes of the economy.

The diffusive term $c(Y(t))dt+\phi(Y(t))dW(t)+\bar{\phi}(Y(t))d\bar{W}(t)$ in \eqref{eq:Ct} models the fluctuations in the value of the claim per policy. From equations~\eqref{eq:tildeP} and \eqref{eq:Ct}, it can be seen that apart from the risk (pure jump) model for the claims, the claim (risk) per policy $C(t)$ is also driven by an idiosyncratic source of risk $\bar{W}$ and has the common source of risk ${W}$ with the defaultable stock prices $\tilde{S}(t)$. Thus, by Zou and Cadenillas~\cite{ZouCa14}, the total risk of the insurer in our case can be described as
\begin{align}\label{eq:risk-model}
dR^\eta(t)=\eta(t)dC(t).
\end{align}
The forthcoming section will formulate the dynamic optimization problem for an insurer and formally derive the recursive HJB system using the dynamic programming principle.

\section{Dynamic Optimization for an Insurer}\label{sec:framework}

In this section, we formulate the optimal investment and risk control problem for an insurer and derive the recursive HJB system accordingly. For this reason,
for $j=1,\ldots,n$, let $\tilde{\pi}_j(t)$ be the $\Gx$-predictable fraction strategy for the $j$-th defaultable stock at time $t$. We assume that the insurer will not invest in the stock once it has defaulted. Then $1-\tilde{\pi}(t)^{\top}e_n^{\top}$ is the fraction strategy for the risk-free money market account at time $t$. The dynamics of the money market account $B(t)$ is given by $dB(t)= r(Y(t))B(t)dt$, where the regime-switching interest rate $r(i)>0$ for $i=1,\ldots,m$. Here $\tilde{\pi}(t):=(\tilde{\pi}_j(t);\ j=1,\ldots,n)^{\top}$ and $e_n$ denotes the $n$-dimensional row vector whose all entities are ones.

We assume that the average premium per liability for the insurer is $p(Y(t),Z(t))$ (i.e., it depends not only on the macro-economy, but also on the default state of the portfolio), then the price of the insurance risk satisfies the dynamics $dP(t)=p(Y(t),Z(t))dt-dC(t)$.
The insurer is in fact able to trade this risk process by selling insurance products and ceding part or all of his/her business to reinsurers.
Recall that $\eta(t)$ stands for the $\Gx$-predictable total outstanding number of policies (liabilities) at time $t$ introduced in Section~\ref{sec:model}.
{Let $X^{\tilde{\pi},\tilde{l}}(t)$ represent the time-$t$ wealth level corresponding to the strategy $(\tilde{\pi},\tilde{l})$,} then the self-financing condition yields that
\begin{align}\label{eq:SDEX}
\frac{dX^{\tilde{\pi},\tilde{l}}(t)}{X^{\tilde{\pi},\tilde{l}}(t-)}&=\tilde{\pi}(t)^{\top}{ diag}(\tilde{S}(t-))^{-1}d\tilde{S}(t)+\big(1-\tilde{\pi}(t)^{\top}e_n^{\top}\big)\frac{dB(t)}{B(t)}+\tilde{l}(t)dP(t)\\
&=\tilde{\pi}(t)^{\top}{ diag}(\tilde{S}(t-))^{-1}d\tilde{S}(t)+\big(1-\tilde{\pi}(t)^{\top}e_n^{\top}\big)\frac{dB(t)}{B(t)}+ p(Y(t),Z(t))\tilde{l}(t)dt-d{R}^{\tilde{l}}(t),\nonumber
\end{align}
where $\tilde{l}(t)$ is the ratio of liabilities over wealth at time $t$. By virtue of the dynamics \eqref{eq:risk-model}, it holds that
\begin{align}\label{eq:risk-model2}
d{R}^{\tilde{l}}(t)=\tilde{l}(t)\left\{c(Y(t))dt+\phi(Y(t))dW(t)+\bar{\phi}(Y(t))d\bar{W}(t)+g(Y(t-))dN(t)\right\}.
\end{align}
Using equations \eqref{eq:SDEX} and \eqref{eq:risk-model2}, the wealth process of the insurer can be rewritten as
\begin{align}\label{eq:sdeX2}
\frac{dX^{\tilde{\pi},\tilde{l}}(t)}{X^{\tilde{\pi},\tilde{l}}(t-)}=&\big[r(Y(t))+\tilde{\pi}(t)^{\top}(\mu(Y(t))-r(Y(t))e_n^{\top})
+\tilde{l}(t)(p(Y(t),Z(t))-c(Y(t)))\big]dt\nonumber\\
&+\big[\tilde{\pi}(t)^{\top}\sigma(Y(t))-\tilde{l}(t)\phi(Y(t))\big]dW(t)-\tilde{l}(t)\bar{\phi}(Y(t))d\bar{W}(t)\\
&-\tilde{\pi}(t)^{\top}dM(t)-\tilde{l}(t)g(Y(t-))dN(t).\nonumber
\end{align}

We next give the definition of the admissible control set which will be used in the paper.
\begin{definition}\label{def:add-con}
The admissible control set $\tilde{\cal U}$ is a class of $\Gx$-predictable feedback strategies $(\tilde{\pi}(t),\tilde{l}(t))_{t\in[0,T]}:=((\tilde{\pi}_j(t);\ j=1,\ldots,n)^{\top},\tilde{l}(t))_{t\in[0,T]}$, given by the Markov control $\tilde{\pi}_j(t):=\pi_j(t,X^{\tilde{\pi},\tilde{l}}(t-),Y(t-),Z(t-))$ for $j=1,\ldots,n$, and the nonnegative Markov control $\tilde{l}(t):=l(t,X^{\tilde{\pi},\tilde{l}}(t-),Y(t-),Z(t-))$ such that the wealth process $X^{\tilde{\pi},\tilde{l}}(t)$ of the insurer is nonnegative for all $t\in[0,T]$. Moreover $\tilde{\pi}_j(t)=\tilde{\pi}_j(t)(1-Z_j(t-))$ for $j=1,\ldots,n$, and the feedback control function $\pi_j$, $j=1,\ldots,n$ and $l$ are assumed to be locally bounded.
We use ${\cal U}$ to denote the set of the above feedback functions $(\pi,l):=((\pi_j; j=1,\ldots,n)^{\top},l)$.
\end{definition}
For $x\in\R_+$, let $U(x):=\frac{1}{\gamma}x^{\gamma}$ with $\gamma\in(0,1)$ be the power (CRRA) utility. We consider the following expected utility maximization problem from terminal wealth of the insurer given by, for $(t,x,i,z)\in[0,T]\times\R_+\times\{1,\ldots,m\}\times{\cal S}$,
\begin{align}\label{eq:value-fcn}
V(t,x,i,z) := \sup_{(\tilde{\pi},\tilde{l})\in\tilde{\cal U}}\Ex\left[U(X^{\tilde{\pi},\tilde{l}}(T))\big|X^{\tilde{\pi},\tilde{l}}(t)=x,Y(t)=i,Z(t)=z\right].
\end{align}
Suppose that $V$ is $C^{1,2}$ in $(t,x)\in[0,T]\times\R_+$ for each $(i,z)\in\{1,\ldots,m\}\times{\cal S}$. Then, It\^o's formula yields that
\begin{align}\label{eq:hjbsystem}
&dV(t,X^{\tilde{\pi},\tilde{l}}(t),Y(t),Z(t))\nonumber\\
&\ =\bigg\{\frac{\partial V}{\partial t}+X^{\tilde{\pi},\tilde{l}}(t)\frac{\partial V}{\partial x}\big[r(Y(t))+\tilde{\pi}(t)^{\top}\theta(Y(t),Z(t))+\tilde{l}(t)(p(Y(t),Z(t))-c(Y(t)))\big]\nonumber\\
&\quad+\frac{1}{2}(X^{\tilde{\pi},\tilde{l}}(t))^2\frac{\partial^2 V}{\partial x^2}\big[\tilde{\pi}(t)^{\top}\sigma(Y(t))\sigma(Y(t))^{\top}\tilde{\pi}(t)
+(\tilde{l}(t))^2(\phi(Y_{t})\phi(Y_{t})^{\top}+\bar{\phi}(Y(t))\bar{\phi}(Y(t))^{\top})\nonumber\\
&\quad\qquad-2\tilde{l}(t)\tilde{\pi}(t)^{\top}\sigma(Y(t))\phi(Y(t))^{\top}\big]\bigg\}dt\nonumber\\
&\quad+X^{\tilde{\pi},\tilde{l}}(t)\frac{\partial V}{\partial x}\left\{\big[\tilde{\pi}(t)^{\top}\sigma(Y(t))-\tilde{l}(t)\phi(Y(t))]dW(t)-\tilde{l}(t)\bar{\phi}(Y(t))d\bar{W}(t)\right\}\nonumber\\
&\quad+\big[V(t,X^{\tilde{\pi},\tilde{l}}(t-)-\tilde{l}(t)X^{\tilde{\pi},\tilde{l}}(t-)g(Y(t-)),Y(t-),Z(t-))-V(t,X^{\tilde{\pi},\tilde{l}}(t-),Y(t-),Z(t-))\big]dN(t)\nonumber\\
&\quad+\sum_{j=1}^n\big[V(t,X^{\tilde{\pi},\tilde{l}}(t-)-\tilde{\pi}_j(t)X^{\tilde{\pi},\tilde{l}}(t-),Y(t-),Z^{j}(t-))-V(t,X^{\tilde{\pi},\tilde{l}}(t-),Y(t-),Z(t-))\big]dZ_j(t)\nonumber\\
&\quad+\sum_{j\neq Y(t-)}[V(t,X^{\tilde{\pi},\tilde{l}}(t-),j,Z(t-))-V(t,X^{\tilde{\pi},\tilde{l}}(t-),Y(t-),Z(t-))\big]dH_{Y(t-),j}(t).
\end{align}
Here, the coefficient $\theta(i,z):=\mu(i)-r(i)e_n^{\top}+h(i,z)$ for $(i,z)\in\{1,\ldots,m\}\times{\cal S}$, and the process, for $t\in[0,T]$,
\begin{align}\label{eq:Hil}
H_{ij}(t):=\sum_{0<s\leq t}\mathds{1}_{Y(t-)=i,Y(t)=j},\quad i,j\in\{1,\ldots,m\}\ {\rm and}\ i\neq j.
\end{align}
The dynamic programming principle yields that the value function $V$ satisfies the following HJB equation, i.e., for $(t,x,i,z)\in[0,T)\times\R_+\times\{1,\ldots,m\}\times{\cal S}$,
\begin{align}\label{eq:hjbsystem2}
0&=\frac{\partial V(t,x,i,z)}{\partial t}+r(i)x\frac{\partial V(t,x,i,z)}{\partial x}+\sum_{j\neq i}\big[V(t,x,j,z)-V(t,x,i,z)\big]q_{ij}\nonumber\\
&\quad+\sup_{(\pi,l)\in{\cal U}}\Bigg\{x\frac{\partial V(t,x,i,z)}{\partial x}\big[\pi^{\top}(I-diag(z))\theta(i,z)+l(p(i,z)-c(i))\big]\nonumber\\
&\quad+\frac{1}{2}x^2\frac{\partial^2 V(t,x,i,z)}{\partial x^2}\big[\pi^{\top}(I-diag(z))\sigma(i)\sigma(i)^{\top}(I-diag(z))\pi+l^2(\phi(i)\phi(i)^{\top}+\bar{\phi}(i)\bar{\phi}(i)^{\top})\nonumber\\
&\quad-2l\pi^{\top}(I-diag(z))\sigma(i)\phi(i)^{\top}\big]\nonumber\\
&\quad+\big[V(t,x-xlg(i),i,z)-V(t,x,i,z)\big]\nu(i,z)\nonumber\\
&\quad+\sum_{j=1}^n\big[V(t,x-\pi_jx,i,z^j)-V(t,x,i,z)\big](1-z_j)h_j(i,z)\Bigg\}
\end{align}
with terminal condition $V(T,x,i,z)=U(x)$ for all $(x,i,z)\in\R_+\times\{1,\ldots,m\}\times{\cal S}$. Here, for $j=1,\ldots,n$, and $z\in{\cal S}$, the flipped state is defined as
\begin{align}\label{eq:zj}
z^j=(z_1,\ldots,z_{j-1},1-z_j,z_{j+1},\ldots,z_n).
\end{align}
In particular, we set $z^j=z$ if $j=0$.

It can be observed that Eq.~\eqref{eq:hjbsystem2} is in fact a recursive dynamical system in terms of default states $z\in{\cal S}$. Further if we consider the value function in the form of $V(t,x,i,z)=x^{\gamma}\varphi(t,i,z)$, then $\varphi(t,i,z)$ satisfies the recursive dynamical system given by, for $(i,z)\in\{1,\ldots,m\}\times{\cal S}$, on $t\in[0,T)$,
\begin{align}\label{eq:hjbeqn}
0&=\frac{\partial \varphi(t,i,z)}{\partial t}+\gamma r(i)\varphi(t,i,z)+\sum_{j=1}^m\varphi(t,j,z)q_{ij}\nonumber\\
&\quad+\sup_{(\pi,l)\in{\cal U}}H\big((\pi,l);i,z,(\varphi(t,i,z^j);\ j=0,1,\ldots,n)\big)
\end{align}
with terminal condition $\varphi(T,i,z)=\frac{1}{\gamma}$ for all $(i,z)\in\{1,\ldots,m\}\times{\cal S}$. For $(\pi,l)\in(-\infty,1]^n\times[0,\infty)$ and $(i,z)\in\{1,\ldots,m\}\times{\cal S}$, the function
\begin{align}\label{eq:H}
&H\big((\pi,l);t,i,z,\bar{f}(z)\big)=\gamma\big\{\pi^{\top}(I-diag(z))\theta(i,z)+(p(i,z)-c(i))l\big\}f(z)\nonumber\\
&\qquad+\bigg\{\frac{\gamma(\gamma-1)}2\pi^{\top}(I-diag(z))\sigma(i)\sigma(i)^{\top}(I-diag(z))\pi+[(1-lg(i))^\gamma-1]\nu(i,z)\nonumber\\
&\qquad+\frac{\gamma(\gamma-1)}2l^2\big(\phi(i)\phi(i)^{\top}+\bar{\phi}(i)\bar{\phi}(i)^{\top}\big)-\gamma(\gamma-1)l\pi^{\top}(I-diag(z))\sigma(i)\phi(i)^{\top}\bigg\}f(z)\nonumber\\
&\qquad+\sum_{j=1}^n[(1-\pi_j)^\gamma f({z}^j)-f(z)](1-z_j)h_j(i,z),
\end{align}
where ${\bar f}(z)=(f(z^j);\ j=0,1,\ldots,n)$ is an arbitrary vector-valued function defined on $z\in{\cal S}$. In the forthcoming section, we will study the existence and uniqueness of (classical) solutions of the recursive HJB system \eqref{eq:hjbeqn}.

\section{Analysis of Iterated HJB Equations}\label{sec:HJB-sys}

This section analyzes the existence and uniqueness of global (classical) solutions to the recursive dynamical system \eqref{eq:hjbeqn} in terms of default states $z\in{\cal S}$.

We introduce the notations which will be used frequently in this section.  For $x\in\mathbb{R}^m$, we write $x=(x_1,...,x_m)^{\top}$ as an $m$-dimensional column vector. For any $x,y\in\R^m$, we write $x\leq y$ if $x_i\leq y_i$ for all $i=1,\ldots,m$, while we write $x<y$ if $x\leq y$ and there exists some $i\in\{1,\ldots,m\}$ such that $x_i<y_i$.
In particular, we write $x\ll y$ if $x_i<y_i$ for all $i=1,2,...,m$. Recall that $e_{n}$ denotes the $n$-dimensional row vector whose all entities are ones.
For the general default state $z\in{\cal S}$, we introduce a general default state representation $z=0^{j_1,\ldots,j_k}$ for indices $j_1\neq\cdots\neq j_k$ belonging to $\{1,\ldots,n\}$, and $k\in\{0,1,\ldots,n\}$. Such a vector $z$ is obtained by flipping the entries $j_1,\ldots,j_k$ of the zero vector to one, i.e., $z_{j_1}=\cdots=z_{j_k}=1$, and $z_{j}=0$ for $j\notin\{j_1,\ldots,j_k\}$ (if $k=0$, we set $z=0^{j_1,\ldots,j_k}=0$). Clearly $0^{j_1,\ldots,j_{n}}=e_n$.

Recall the recursive dynamical system \eqref{eq:hjbeqn} in terms of default states $z=0^{j_1,\ldots,j_k}$ (where $k=0,1,\ldots,n$). The solvability can in fact be  analyzed in the recursive form in terms of default states. Hence, our proof strategy for analyzing the system is based on a recursive procedure, starting from the default state $z=e_n$ (i.e., all stocks have defaulted) and proceeding backward to the default state $z=0$ (i.e., all stocks are alive).
\begin{itemize}
  \item[(i)] $k=n$ (i.e., all stocks have defaulted). In this default state, the insurer will not invest in stocks because they have defaulted and hence the optimal fraction strategy for stocks is given by $\pi_1^*=\cdots=\pi_n^*=0$ by virtue of Definition~\ref{def:add-con}. Let $\varphi(t,e_n)=(\varphi(t,i,e_n);\ i=1,\ldots,m)^{\top}$. Then, the dynamical system \eqref{eq:hjbeqn} reduces to
      \begin{align}\label{eq:hjben}
\left\{
\begin{aligned}
\frac d{dt}\varphi(t,e_n)=&-A^{(n)}\varphi(t,e_n),\quad\text{ in }[0,T);\\
\varphi(T,e_n)=&\frac{1}{\gamma}e_m^{\top}.
\end{aligned}
\right.
\end{align}
\end{itemize}
Here the matrix of coefficient is given by
\begin{align}\label{eq:Aen}
A^{(n)}=&diag\left[\left(\gamma r(i)+\sup_{l\in{\cal U}^{(n)}}H^{(n)}(l,i);\ i=1,\ldots,m\right)\right]+Q,
\end{align}
where the policy space in \eqref{eq:Aen} in this case is reduced to
\begin{align}\label{eq:Un}
{\cal U}^{(n)}:=\{l=l(i)\in[0,+\infty);\ 1-lg(i)\geq0\}.
\end{align}
Moreover, the function $H^{(n)}(l,i)$ is given by, for $(l,i)\in[0,\infty)\times\{1,\ldots,m\}$,
\begin{align*}
H^{(n)}(l,i):=&\gamma(p(i,e_n)-c(i))l+\frac{\gamma(\gamma-1)}{2}l^2\big(\phi(i)\phi(i)^{\top}+\bar{\phi}(i)\bar{\phi}(i)^{\top}\big)\\
&+[(1-lg(i))^\gamma-1]\nu(i,e_n).
\end{align*}
Since $\gamma\in(0,1)$, it is not difficult to verify that for each $i=1,\ldots,m$, $H^{(n)}(l,i)$ is continuous and strictly concave in $l$ on the compact ${\cal U}^{(n)}$. Consequently, there exists a unique optimum $l^{*}\in{\cal U}^{(n)}$ which is given by
\begin{align}\label{eq:lnstar}
l^{*} =l^{*}(i)= \argmax_{l\in{\cal U}^{(n)}}H^{(n)}(l,i),\qquad i=1,\ldots,m.
\end{align}
Further, we have that $\sup_{l\in{\cal U}^{(n)}}H^{(n)}(l,i)=H^{(n)}(l^{*},i)\in[0,\infty)$ for each $i=1,\ldots,m$. Then, the matrix of coefficient $A^{(n)}$ given by \eqref{eq:Aen} is finite.

We next prove that the dynamical system \eqref{eq:hjben} has a unique strictly positive solution. To this purpose, we need the following auxiliary result which will be also used in the proof related to the general default case. The proof is provided in the Appendix.
\begin{lemma}\label{lem:sol-hjben2}
Let $g(t):=(g_i(t);\ i=1,\ldots,m)^{\top}$ satisfy the following dynamical system given by
\begin{align*}
\left\{
\begin{aligned}
\frac d{dt}g(t)=&Bg(t)\quad\text{ in }(0,T];\\
g(0)=&\xi.
\end{aligned}
\right.
\end{align*}
If $B=(b_{ij})_{m\times m}$ satisfies $b_{ij}\geq 0$ for $i\neq j$ and $\xi\gg0$,
then $g(t)\gg0$ for all $t\in[0,T]$.
\end{lemma}
Then we have the following lemma whose proof is given in the Appendix.
\begin{lemma}\label{lem:sol-hjben}
The dynamical system \eqref{eq:hjben} admits a unique solution which is given by, for $t\in[0,T]$,
\begin{align}\label{eq:varphien}
\varphi(t,e_n)=\frac{1}{\gamma} e^{A^{(n)}(T-t)}e_m^{\top}=\frac{1}{\gamma}\sum_{i=0}^{\infty}\frac{(A^{(n)})^i(T-t)^i}{i!}e_m^{\top},
\end{align}
where the $m\times m$-dimensional matrix $A^{(n)}$ is given by \eqref{eq:Aen}. Moreover, it holds that $\varphi(t,e_n)\gg 0$ for all $t\in[0,T]$.
\end{lemma}

We next consider the general default state with the form $z=0^{j_1,\ldots,j_{k}}$ for $0\leq k\leq n-1$, i.e., the stocks $j_1,\ldots,j_{k}$ have defaulted and the stocks $\{j_{k+1},\ldots,j_n\}:=\{1,\ldots,n\}\setminus\{j_1,\ldots,j_k\}$ are alive. Then, we have
\begin{itemize}
  \item[(ii)] Since the stocks $j_1,\ldots,j_k$ have defaulted, the optimal fraction strategies for the stocks $j_1,\ldots,j_{k}$ are given by $\pi_j^{(k,*)}=0$ for $j\in\{j_1,\ldots,j_{k}\}$ by virtue of Definition~\ref{def:add-con}. Let $\varphi^{(k)}(t)=(\varphi(t,i,0^{j_1,\ldots,j_{k}});\ i=1,\ldots,m)^{\top}$, $p^{(k)}(i)=p(i,0^{j_1,\ldots,j_{k}})$, and $h^{(k)}_j(i)=h_j(i,0^{j_1,\ldots,j_{k}})$ for $j\notin\{j_1,\ldots,j_k\}$ and $i=1,\ldots,m$. Therefore, the corresponding HJB system \eqref{eq:hjbeqn} in this default state reduces to
       \begin{align}\label{eq:hjbn-1}
\left\{
\begin{aligned}
\frac d{dt}\varphi^{(k)}(t)=&-A^{(k)}\varphi^{(k)}(t)-G^{(k)}(t,\varphi^{(k)}(t)),\quad\text{ in }[0,T);\\
\varphi^{(k)}(T)=&\frac{1}{\gamma}e_m^{\top}.
\end{aligned}
\right.
\end{align}
Here, the $m\times m$-dimensional matrix $A^{(k)}$ is given by
\begin{align}\label{eq:An-1}
A^{(k)}=diag\left[\left(\gamma r(i)-\sum_{j\notin\{j_1,\ldots,j_{k}\}}h_{j}^{(k)}(i);\ i=1,\ldots,m\right)\right]+Q.
\end{align}
The coefficient $G^{(k)}(t,x)=(G^{(k)}_i(t,x);\ i=1,\ldots,m)^{\top}$ for $(t,x)\in[0,T]\times\R^{m}$ is given by, for $i=1,\ldots,m$,
\begin{align}\label{eq:Gin-1}
G_i^{(k)}(t,x)&=\sup_{(\pi^{(k)},l)\in{\cal U}^{(k)}}\left\{\sum_{j\notin\{j_1,\ldots,j_k\}}(1-\pi_{j}^{(k)})^\gamma h^{(k)}_{j}(i)\varphi^{(k+1),j}(t,i)+H^{(k)}((\pi^{(k)},l),i)x_i\right\},
\end{align}
where the policy space in this default case is given by
\begin{align}\label{eq:Un}
{\cal U}^{(k)}:=\left\{(\pi^{(k)},l)=(\pi^{(k)}(t,i),l(t,i))\in(-\infty,1]^{n-k}\times[0,\infty);\ 1-lg(i)\geq0\right\}.
\end{align}
The function $\varphi^{(k+1),j}(t,i):=\varphi(t,i,0^{j_1,\ldots,j_k,j})$ for $j\notin\{j_1,\ldots,j_k\}$ corresponds to the $i$-th element of the positive solution of the HJB system \eqref{eq:hjbeqn} at the default state $z=0^{j_1,\ldots,j_k,j}$. The function $H^{(k)}((\pi^{(k)},l),i)$ is given by, for $(\pi^{(k)},l)\in{\cal U}^{(k)}$, and $i=1,\ldots,m$,
\begin{align}\label{eq:Hk}
H^{(k)}((\pi^{(k)},l),i)=&\gamma\big\{(\pi^{(k)})^{\top}\theta^{(k)}(i)+(p^{(k)}(i)-c(i))l\big\}+\frac{\gamma(\gamma-1)}2\Big\{(\pi^{(k)})^{\top}\sigma^{(k)}(i)\sigma^{(k)}(i)^{\top}\pi^{(k)}\nonumber\\
&+l^2\big[\phi(i)\phi(i)^{\top}+\bar{\phi}(i)\bar{\phi}(i)^{\top}\big]-2l(\pi^{(k)})^{\top}\sigma^{(k)}(i)\phi(i)^{\top}\Big\}\nonumber\\
&+\big[(1-lg(i))^\gamma-1\big]\nu^{(k)}(i).
\end{align}
Here, for each $i=1,\ldots,m$, we used notations $\pi^{(k)}=(\pi_j^{(k)};\ j\notin\{j_1,\ldots,j_k\})^{\top}$, $\theta^{(k)}(i)=(\theta_j(i);\ j\notin\{j_1,\ldots,j_k\})^{\top}$, $\sigma^{(k)}(i)=(\sigma_{j\kappa}(i);\ j\notin\{j_1,\ldots,j_k\},\kappa\in\{1,\ldots,d\})$, and $\nu^{(k)}(i)=\nu(i,0^{j_1,\ldots,j_k})$.
\end{itemize}
From the expression of $G_i^{(k)}(t,x)$ given by \eqref{eq:Gin-1}, it can be seen that the solution $\varphi^{(k)}(t)$ on $t\in[0,T]$ of Eq.~\eqref{eq:hjbeqn} at $z=0^{j_1,\ldots,j_k}$ depends on the solution $\varphi^{(k+1),j}(t)$ on $t\in[0,T]$ of Eq.~\eqref{eq:hjbeqn} at
$z=0^{j_1,\ldots,j_k,j}$ for $j\notin\{j_1,\ldots,j_k\}$. In particular, for $k=n-1$, the solution $\varphi^{(k+1),j}(t)=\varphi(t,e_n)\gg0$ corresponding to the solution to Eq.~\eqref{eq:hjbeqn} at $z=e_n$ (i.e., $k=n$) has been obtained in Lemma~\ref{lem:sol-hjben}. This suggests solving the HJB system \eqref{eq:hjbeqn} backward recursively in terms of default states $z=0^{j_1,\ldots,j_k}$. Thus, in order to analyze the existence and uniqueness of a positive (classical) solution to the dynamical system \eqref{eq:hjbn-1}, we first assume that the HJB system \eqref{eq:hjbeqn} admits a positive unique (classical) solution $\varphi^{(k+1),j}(t)$ on $t\in[0,T]$ for $j\notin\{j_1,\ldots,j_k\}$.

We have the following estimate on $G^{(k)}(t,x)$ given by \eqref{eq:Gin-1} which is stated in the following lemma. The proof is reported in the Appendix.
\begin{lemma}\label{lem:Gkesti}
For each $k=0,1,\ldots,n-1$, assume that the HJB system \eqref{eq:hjbeqn} admits a positive unique (classical) solution $\varphi^{(k+1),j}(t)$ on $t\in[0,T]$ for $j\notin\{j_1,\ldots,j_k\}$. Then, for any $x,y\in\R^m$ satisfying $x,y\geq\varepsilon e_m^{\top}$ with $\varepsilon>0$, there exists a positive constant $C=C(\varepsilon)$ depending on $\varepsilon>0$ only such that
\begin{align}\label{eq:Gkesti}
\left\|G^{(k)}(t,x)-G^{(k)}(t,y)\right\|\leq C\left\|x-y\right\|.
\end{align}
Here $\|\cdot\|$ denotes the Euclidian norm.
\end{lemma}

In order to study the existence and uniqueness of solutions to the HJB system \eqref{eq:hjbn-1}, we also need the following comparison result. The proof is delegated to the Appendix.
\begin{lemma}\label{lem:comparison}
Let $g_{\kappa}(t):=(g_{\kappa i}(t);\ i=1,\ldots,m)^{\top}$ with $\kappa=1,2$ satisfy the following dynamical systems on $[0,T]$ respectively
\begin{align*}
\left\{
\begin{aligned}
\frac d{dt}g_1(t)=&f(t,g_1(t))+\tilde{f}(t,g_1(t)),\ \text{ in }(0,T];\\
g_1(0)=&\xi_1,
\end{aligned}
\right.\quad\quad
\left\{
\begin{aligned}
\frac d{dt}g_2(t)=&f(t,g_2(t)),\ \text{ in }(0,T];\\
g_2(0)=&\xi_2.
\end{aligned}
\right.
\end{align*}
Here the functions $f(t,x),\,\tilde{f}(t,x):[0,T]\times\R^m\to\R^m$ are Lipschitz continuous w.r.t. $x\in\R^m$ uniformly in $t\in[0,T]$. The function
$f(t,\cdot)$ satisfies the type $K$ condition for each $t\in[0,T]$ (i.e., for any $x,y\in\R^m$ satisfying $x\leq y$ and $x_i=y_i$ for some $i=1,\ldots,m$,
it holds that $f_i(t,x)\leq f_i(t,y)$ for each $t\in[0,T]$). If
$\tilde{f}(t,x)\geq0$ for $(t,x)\in[0,T]\times\R^m$ and $\xi_1\geq\xi_2$, then $g_1(t)\geq g_2(t)$ for all $t\in[0,T]$.
\end{lemma}

We are now at the position to state the result of existence and uniqueness of positive (classical) solutions to the HJB system \eqref{eq:hjbn-1}.
\begin{theorem}\label{thm:solutionk}
For each $k=0,1,\ldots,n-1$, assume that the HJB system \eqref{eq:hjbeqn} admits a positive unique (classical) solution $\varphi^{(k+1),j}(t)$ on $t\in[0,T]$ for $j\notin\{j_1,\ldots,j_k\}$. Then, there exists a unique positive (classical) solution $\varphi^{(k)}(t)$ on $t\in[0,T]$ of the HJB system \eqref{eq:hjbeqn}
at the default state $z=0^{j_1,\ldots,j_k}$ (i.e., the HJB system \eqref{eq:hjbn-1} admits a unique positive (classical) solution).
\end{theorem}

\noindent{\it Proof.}\quad For a constant $a>0$, consider the following truncated dynamical system given by
\begin{align}\label{eq:truneq1}
\left\{
\begin{aligned}
\frac{d}{dt}\varphi_a^{(k)}(t)=&-A^{(k)}\varphi^{(k)}_a(t)-G^{(k)}_{a}(t,\varphi_a^{(k)}(t)),\ \text{ in }[0,T);\\
\varphi_a^{(k)}(T)=&\frac{1}{\gamma}e_m^{\top},
\end{aligned}
\right.
\end{align}
where the truncated nonlinearity $G_a^{(k)}(t,x):=G^{(k)}(t,x\vee ae_m^{\top})$ for $(t,x)\in[0,T]\times\R^m$. Thanks to Lemma~\ref{lem:Gkesti}, there exists a positive constant $C=C(a)$ which depends on $a>0$ only such that for all $t\in[0,T]$,
\begin{align}\label{eq:Lip-Ga}
\big\|G_a^{(k)}(t,x)-G_a^{(k)}(t,y)\big\|\leq C\|x-y\|,\qquad x,y,\in\R^m,
\end{align}
i.e., $G^{(k)}_a(t,x)$ is globally Lipschitz continuous w.r.t. $x\in\R^m$ uniformly in $t\in[0,T]$. By reversing the flow of time, consider $\tilde{\varphi}_a^{(k)}(t):=\varphi_a^{(k)}(T-t)$ for $t\in[0,T]$. Then $\tilde{\varphi}_a^{(k)}(t)$ satisfies the following dynamical system given by
\begin{align}\label{eq:truneq2}
\left\{
\begin{aligned}
\frac{d}{dt}\tilde{\varphi}_a^{(k)}(t)=&A^{(k)}\tilde{\varphi}^{(k)}_a(t)+G^{(k)}_{a}(T-t,\tilde{\varphi}_a^{(k)}(t)),\ \text{ in }(0,T];\\
\tilde{\varphi}_a^{(k)}(0)=&\frac{1}{\gamma}e_m^{\top}.
\end{aligned}
\right.
\end{align}
Let $\psi^{(k)}(t)=(\psi_i^{(k)}(t);\ i=1,\ldots,m)^{\top}$ satisfy the following dynamical system:
\begin{align}\label{eq:psieqn}
\left\{
\begin{aligned}
\frac d{dt}\psi^{(k)}(t)=&A^{(k)}\psi^{(k)}(t),\quad\text{ in }(0,T];\\
\psi^{(k)}(0)=&\frac{1}{\gamma}e_m^{\top}.
\end{aligned}
\right.
\end{align}
Recall the $m\times m$-dimensional matrix of coefficients $A^{(k)}$ given by \eqref{eq:An-1}.  Then, we have that $[A^{(k)}]_{ij}=q_{ij}$ for all $i\neq j$ using \eqref{eq:An-1}. Since~$Q=(q_{ij})_{m\times m}$~is the generator of the Markov chain, it holds that $q_{ij}\geq0$ for all $i\neq j$. Hence $[A^{(k)}]_{ij}\geq0$ for all $i\neq j$ and thus the linear function $A^{(k)}x$ is of type $K$ in $x\in\R^m$. Also since $\psi^{(k)}(0)=\frac{1}{\gamma}e_m^{\top}\gg0$, it follows from Lemma~\ref{lem:sol-hjben2} that the dynamical system \eqref{eq:psieqn} admits a unique (classical) solution $\psi^{(k)}(t)$ on $[0,T]$ and moreover $\psi^{(k)}(t)\gg0$ for all $t\in[0,T]$. Set
\begin{align}\label{eq:epsilonk}
\varepsilon^{(k)}:=\min_{i=1,\ldots,m}\left\{\inf_{t\in[0,T]}\psi_i^{(k)}(t)\right\}.
\end{align}
Then, by the continuity of $\psi^{(k)}(t)$ in $t\in[0,T]$ and the fact that $\psi^{(k)}(t)\gg0$ for all $t\in[0,T]$, we have that $\varepsilon^{(k)}>0$. Further, by virtue of estimates \eqref{eq:Lip-Ga} and \eqref{eq:G>0} in the Appendix, together with the initial condition $\varphi_a^{(k)}(0)=\psi^{(k)}(0)=\frac{1}{\gamma}e_m^{\top}\gg0$, it follows from Lemma~\ref{lem:comparison} that \begin{align}\label{eq:tildesol}
\tilde{\varphi}_a^{(k)}(t)\geq \psi^{(k)}(t)\geq \varepsilon^{(k)}e_m^{\top},\quad \text{for all}\ t\in[0,T].
\end{align}
Notice that the positive constant $\varepsilon^{(k)}$ is independent of $a>0$. Then, for $a\in(0,\varepsilon^{(k)})$, it holds that
\begin{align*}
G_a^{(k)}(T-t,\tilde{\varphi}^{(k)}_a(t))=G^{(k)}\big(T-t,\tilde{\varphi}^{(k)}_a(t)\vee ae_m^{\top}\big)=G^{(k)}(T-t,\tilde{\varphi}^{(k)}_a(t)).
\end{align*}
This yields that for $a\in(0,\varepsilon^{(k)})$, the function $\tilde{\varphi}^{(k)}_a(t)$ solves the dynamical system given by
\begin{align*}
\left\{
\begin{aligned}
\frac{d}{dt}\tilde{\varphi}_a^{(k)}(t)=&A^{(k)}\tilde{\varphi}^{(k)}_a(t)+G^{(k)}(T-t,\tilde{\varphi}_a^{(k)}(t)),\ \text{ in }(0,T];\\
\tilde{\varphi}_a^{(k)}(0)=&\frac{1}{\gamma}e_m^{\top}.
\end{aligned}
\right.
\end{align*}
By the uniqueness of the solution to the dynamical system \eqref{eq:truneq2} and using the estimate \eqref{eq:tildesol}, it follows that, for $a\in(0,\varepsilon^{(k)})$, $\varphi_a^{(k)}(t):=\tilde{\varphi}_a^{(k)}(T-t)$ on $[0,T]$ is the unique (classical) solution to the HJB system \eqref{eq:hjbn-1}. Thus, we complete the proof of the theorem. \hfill$\Box$\\

We next turn to the characterization of the optimal strategy $(\pi^{(k)},l)\in{\cal U}^{(k)}$ at the default state $z=0^{j_1,\ldots,j_k}$ where $k=0,1,\ldots,n-1$. Let us recall the HJB system \eqref{eq:hjbn-1}, i.e.,
\begin{align*}
\left\{
\begin{aligned}
\frac d{dt}\varphi^{(k)}(t)=&-A^{(k)}\varphi^{(k)}(t)-G^{(k)}(t,\varphi^{(k)}(t)),\quad\text{ in }[0,T);\\
\varphi^{(k)}(T)=&\frac{1}{\gamma}e_m^{\top}.
\end{aligned}
\right.
\end{align*}
Theorem~\ref{thm:solutionk} shows that the above system admits a unique positive (classical) solution $\varphi^{(k)}(t)$ on $[0,T]$ and moreover $\varphi^{(k)}(t)\geq \varepsilon^{(k)}e_m^{\top}$ for all $t\in[0,T]$ using \eqref{eq:tildesol}. Here $\varepsilon^{(k)}>0$ is given by~\eqref{eq:epsilonk}. Then, by virtue of the equality \eqref{eq:Gk2} given in the Appendix, there exists a positive constant $C(\varepsilon^{(k)})$ depending on $\varepsilon^{(k)}>0$ such that for each $i=1,\ldots,m$,
\begin{align}\label{eq:Gk3}
&G^{(k)}_i(t,\varphi^{(k)}(t,i))\\
&\quad=\sup_{\substack{(\pi^{(k)},l)\in{\cal U}^{(k)}\\\|\pi^{(k)}\|^2+l^2\leq C(\varepsilon^{(k)})}}\left\{\sum_{j\notin\{j_1,\ldots,j_k\}}(1-\pi_{j}^{(k)})^\gamma h_{j}^{(k)}(i)\varphi^{(k+1),j}(t,i)+H^{(k)}((\pi^{(k)},l),i)\varphi^{(k)}(t,i)\right\}.\nonumber
\end{align}
Here, for each $i=1,\ldots,m$, $\varphi^{(k+1),j}(t,i)$ on $t\in[0,T]$ is the $i$-th element of the positive (classical) solution $\varphi^{(k+1),j}(t)$ of the HJB system \eqref{eq:hjbeqn} at the default state $z=0^{j_1,\ldots,j_k,j}$ for $j\notin\{j_1,\ldots,j_k\}$. It is not difficult to verify that, for each $i=1,\ldots,m$ and fixed $t\in[0,T]$,
\[
\sum_{j\notin\{j_1,\ldots,j_k\}}(1-\pi_{j}^{(k)})^\gamma h_{j}^{(k)}(i)\varphi^{(k+1),j}(t,i)+H^{(k)}((\pi^{(k)},l),i)\varphi^{(k)}(t,i)
\]
is strictly concave in $(\pi^{(k)},l)\in{\cal U}^{(k)}$. Also notice that the space ${\cal U}^{(k)}\cap\{(\pi^{(k)},l);\ \|\pi^{(k)}\|^2+|l|^2\leq  C(\varepsilon^{(k)})\}$ is compact. Hence, there exists a unique optimum $(\pi^{(k,*)},l^*)\in{\cal U}^{(k)}$ such that
\begin{align}\label{eq:optimumk}
&(\pi^{(k,*)},l^*)=(\pi^{(k,*)}(t,i),l^*(t,i))\\
&\quad=\argmax_{\substack{(\pi^{(k)},l)\in{\cal U}^{(k)}\\\|\pi^{(k)}\|^2+l^2\leq C(\varepsilon^{(k)})}}\left\{\sum_{j\notin\{j_1,\ldots,j_k\}}(1-\pi_{j}^{(k)})^\gamma h_{j}^{(k)}(i)\varphi^{(k+1),j}(t,i)+H^{(k)}((\pi^{(k)},l),i)\varphi^{(k)}(t,i)\right\}\nonumber\\
&\quad=\argmax_{(\pi^{(k)},l)\in{\cal U}^{(k)}}\left\{\sum_{j\notin\{j_1,\ldots,j_k\}}(1-\pi_{j}^{(k)})^\gamma h_{j}^{(k)}(i)\varphi^{(k+1),j}(t,i)+H^{(k)}((\pi^{(k)},l),i)\varphi^{(k)}(t,i)\right\}\nonumber
\end{align}
for all $i=1,\ldots,m$.

We conclude this section with a verification theorem whose proof is reported in the Appendix.
\begin{theorem}\label{thm:vefi}
At any default state $z=0^{j_1,\ldots,j_k}$ for $k=0,1,\ldots,n$, let $\varphi(t,z)$ be the unique positive (classical) solution to the dynamical system of HJB equations \eqref{eq:hjbeqn} (i.e., for $k=n$, $\varphi(t,z)=\varphi(t,e_n)$ is given in Lemma~\ref{lem:sol-hjben} and for $k=0,1,\ldots,n-1$, $\varphi(t,z)=\varphi^{(k)}(t)$ is given in Theorem~\ref{thm:solutionk}). Also let the optimal strategy $(\pi^{*},l^*)=(\pi^*(t,i,z),l^*(t,i,z))$ for $i=1,\ldots,m$ be given by \eqref{eq:lnstar} for $k=n$ and given by \eqref{eq:optimumk} for $k=0,1,\ldots,n-1$. Then, we have that
\begin{itemize}
  \item[(i)] For $(t,x,i,z)\in[0,T]\times\R_+\times\{1,\ldots,m\}\times{\cal S}$, and any admissible feedback strategy $(\pi,l)\in{\cal U}$, it holds that
\begin{align*}
x^\gamma \varphi(t,i,z)\geq \Ex\big[U(X^{\pi,l}(T))\mid X^{{\pi},l}(t)=x,Y(t)=i,Z(t)=z\big].
\end{align*}
\item[(ii)] The value function $V(t,x,i,z)$ for $(t,x,i,z)\in[0,T]\times\R_+\times\{1,\ldots,m\}\times{\cal S}$ admits the following representation
\begin{align*}
V(t,x,i,z)=\Ex\big[U(X^{\pi^*,l^*}(T))| X^{\pi^*,l^*}(t)=x,Y(t)=i,Z(t)=z]=x^{\gamma}\varphi(t,i,z).
\end{align*}
\end{itemize}
\end{theorem}

\section{Numerical Analysis}\label{sec:numerics}

In this section, we investigate the sensitivity of the optimal strategy of stocks and risk control to changes in market parameters. The sensitivity analysis is performed on a simple market model consisting of two defaultable stocks and a riskless bond, i.e., $n=2$. In this market model, it follows from \eqref{eq:P} that the pre-default prices of stocks are given by
\begin{align*}
\left\{
  \begin{array}{ll}
    \frac{dS_1(t)}{S_1(t)}=\{\mu_1(Y(t))+h_1(Y(t),Z(t))\}dt+\sum_{j=1}^2\sigma_{1j}(Y(t))dW_j(t);\\ \\
    \frac{dS_2(t)}{S_2(t)}=\{\mu_2(Y(t))+h_2(Y(t),Z(t))\}dt+\sum_{j=1}^2\sigma_{2j}(Y(t))dW_j(t),
  \end{array}
\right.
\end{align*}
where $Z:=(Z_1,Z_2)\in{\cal S}=\{0,1\}^2$ is the two-dimensional default state process of stocks and $W$ is a two-dimensional Brownian motion (i.e., $d=2$). The regime-switching process $Y$ is a continuous-time (conservative) Markov chain with state space $\{1,2\}$ (i.e., $m=2$). The claim (risk) per policy in the risk control is then given by
\begin{align*}
dC(t) &= c(Y(t))dt+\sum_{j=1}^2\phi_j(Y(t))dW_j(t) + \bar{\phi}(Y(t))d\bar{W}(t)\nonumber\\
&\quad+\sum_{(i,z)\in\{1,2\}\times\{0,1\}^2}g(Y(t-))\mathds{1}_{Y(t-)=i,Z(t-)=z}dN_{i,z}(t).
\end{align*}
Here $\bar{W}$ is a scalar Brownian motion (i.e., $\bar{d}=1$) and $N_{i,z}$ for $(i,z)\in\{1,2\}\times\{0,1\}^2$ are independent Poisson processes with respective intensities $\nu^{z}(i):=\nu(i,z)$. Throughout the section, we use the following benchmark parameters given in Table~\ref{tab:parameters}. In particular, we use the notation $h_k^{z}:=(h_k(1,z),h_k(2,z))$ to represent the vector of default intensities of the $k$-th stock at the default state $z\in\{0,1\}^2$.
\begin{table}[htbp]
\centering
 \begin{tabular}{|c|c|c|c|c|c|c|c|}
 \hline
 \hline
 $\mu(1)$&$\mu(2)$&$r(1)$&$r(2)$&$p(1)$&$p(2)$&$c(1)$&$c(2)$\\
 \hline
 $(1,0.55)$&$(1.4,0.8)$&0.1&0.06&$0.8$&$0.5$&$0.1$&$0.05$\\
 \hline
 \hline
 $\bar{\phi}(1)$&$\bar{\phi}(2)$&$g(1)$&$g(2)$&$h^{(1,0)}_2$&$h^{(0,1)}_1$&$h^{(0,0)}_1$&$h^{(0,0)}_2$\\
 \hline
  0.3&0.6&0.2&0.1&$(0.9,1.3)$&$(0.7,1)$&$(0.5,0.75)$&$(0.75,1.1)$\\
 \hline
 \hline
 $\phi(1)$&$\phi(2)$ &$\nu^{(0,0)}(1)$ & $\nu^{(0,0)}(2)$&$\nu^{(1,0)}(1)$&$\nu^{(1,0)}(2)$&$\nu^{(0,1)}(1)$&$\nu^{(0,1)}(2)$\\
 \hline
 $(0.4,0.8)$&$(0.7,1.2)$ &2 & 3&2.5 &4 &2.3 &3.7\\
 \hline
 \hline
 $\nu^{(1,1)}(1)$&$\nu^{(1,1)}(2)$& & & & & & \\
 \hline
 2.6&5 & & & & & &\\
 \hline
 \end{tabular}
\caption{Market parameters values}\label{tab:parameters}
\end{table}
Moreover, we set the risk aversion parameter to $\gamma=0.5$. The generator of the Markov chain $Y$ and the volatility matrix of stocks are given respectively by
\begin{align*}
Q=Q_0=\left[\begin{matrix}
-0.5&0.5\\
1&-1\\
\end{matrix}\right],\quad \sigma(1)=\left[\begin{matrix}
0.7&0\\
0&1\\
\end{matrix}\right],\quad \sigma(2)=\left[\begin{matrix}
1&0\\
0&1.5\\
\end{matrix}\right].
\end{align*}

\begin{center}
\makeatletter
\def\@captype{figure}
\makeatother
\begin{figure}[htb!]
\hspace*{-10mm}
\begin{minipage}[t]{0.25\linewidth}
\centering
\includegraphics[width=3.3in]{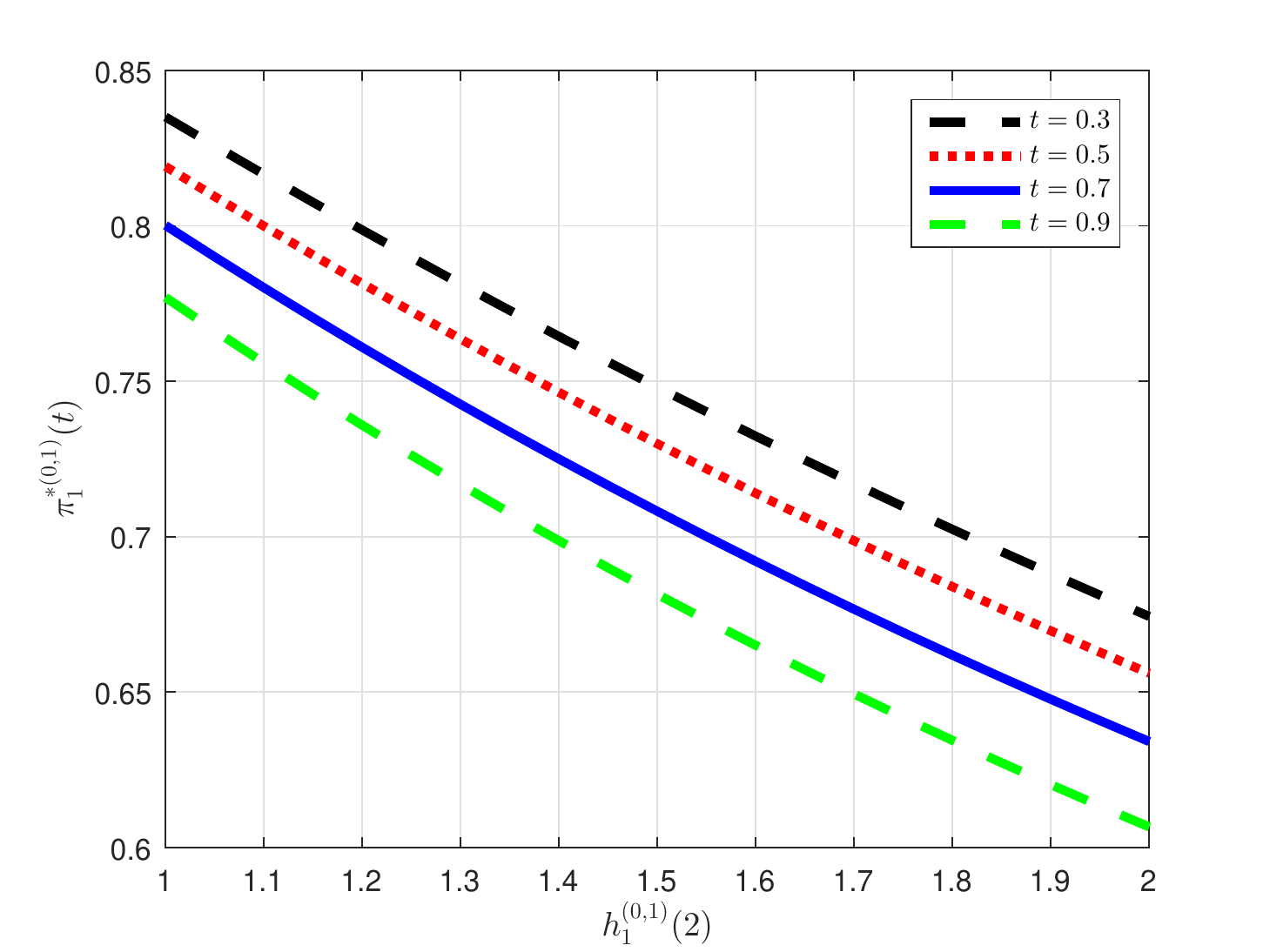}
\end{minipage}
\hspace{28ex}
\begin{minipage}[t]{0.25\linewidth}
\centering
\includegraphics[width=3.3in]{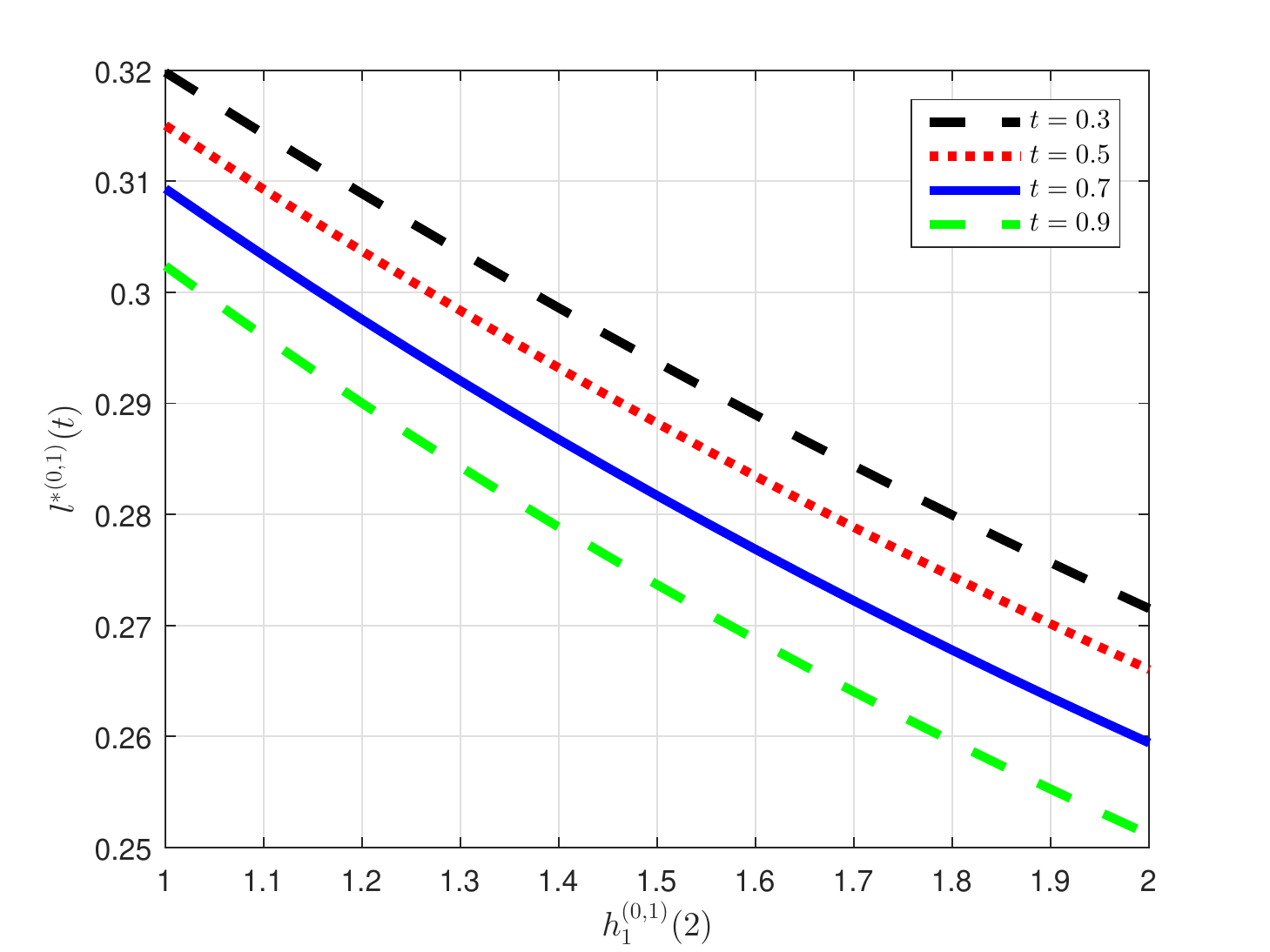}
\end{minipage}

\hspace*{-10mm}
\begin{minipage}[t]{0.25\linewidth}
\centering
\includegraphics[width=3.3in]{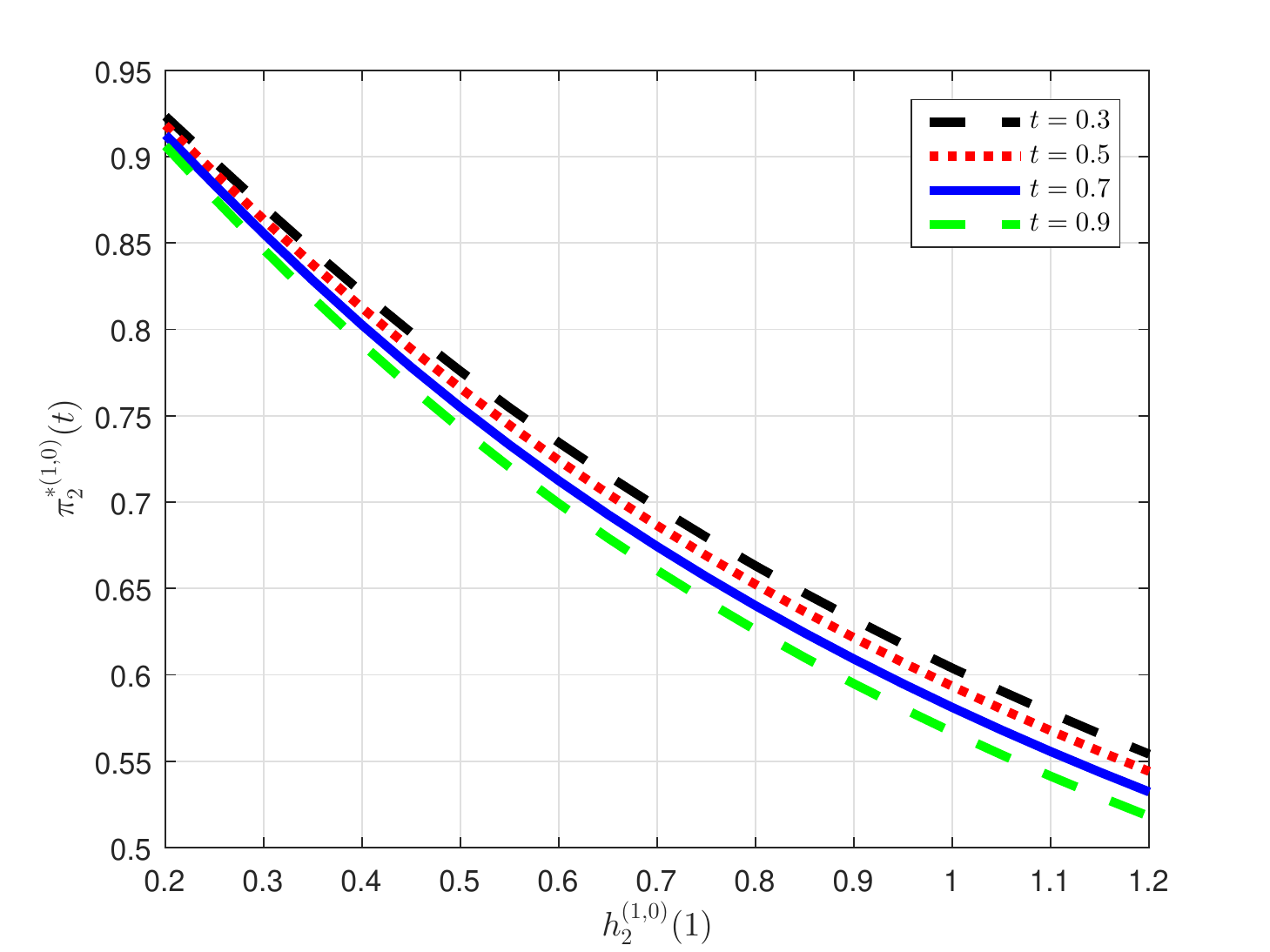}
\end{minipage}
\hspace{28ex}
\begin{minipage}[t]{0.25\linewidth}
\centering
\includegraphics[width=3.3in]{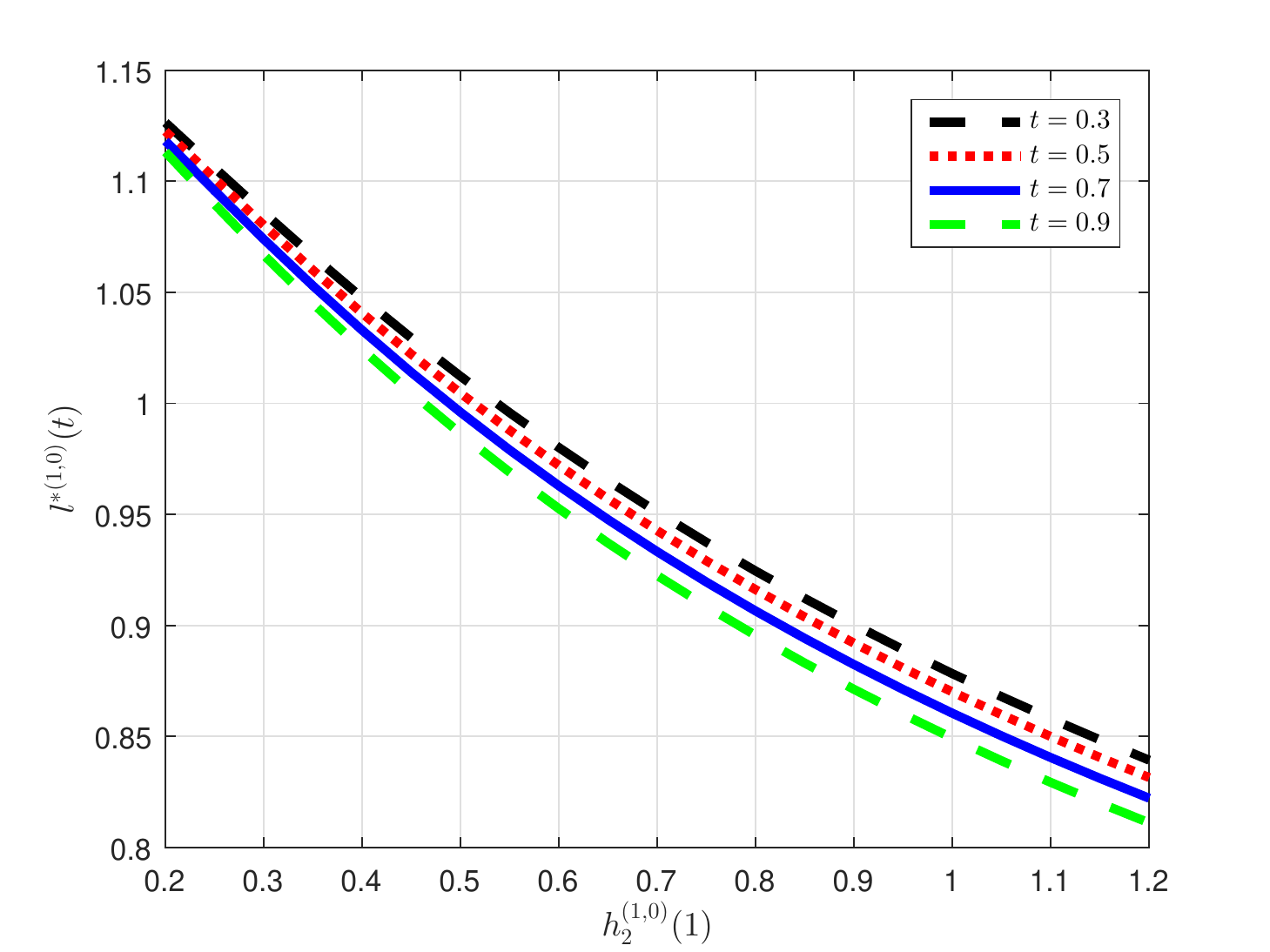}
\end{minipage}
\caption{\small Dependence of the optimal strategies of stocks and risk control on default intensities at a given regime. Top panel: the dependence of the optimal strategies of stock 1 and risk control on the default intensity of stock 1 in regime 2. The default state $z=(0,1)$. Bottom panel: the dependence of the optimal strategies of stock 2 and risk control on the default intensity of stock 2 in regime 1. The default state $z=(1,0)$.}\label{fig:l pi to h}
\end{figure}
\end{center}
We first perform a comparative statics analysis to examine how the default risk premia affect the
optimal strategies of stocks and risk control of the insurer. Figure \ref{fig:l pi to h} displays the optimal strategy of stocks and risk control in a given regime at different times when the default intensity of a stock varies. Consider first the situation in which stock 1 is alive and stock 2 has defaulted (i.e., it corresponds to the default state $z=(0,1)$). The top left graph of Figure~\ref{fig:l pi to h} indicates that, as the stock 1's default intensity becomes higher in regime 2, i.e., $h_1^{(0,1)}(2)$ increases, the insurer reduces his/her investment in the defaultable stock 1. Recall that, for a fixed regime $i\in\{1,2\}$, $\nu^{z}(i)$ represents the jump intensity of the claim (risk) per policy in the risk control at the default state $z\in\{0,1\}^2$. Under the benchmark parameter configuration, we have $\nu^{(1,1)}(2)>\nu^{(0,1)}(2)$ and $\nu^{(1,1)}(1)>\nu^{(1,0)}(1)$. This implies that a default event can result in an increase in the expected number of claims that occur during a fixed period of time. In other words, when the default intensity of a stock increases,
not only the defaultable stocks but the liabilities become riskier. The top right graph of Figure~\ref{fig:l pi to h} shows that, as the default intensity of stock 1 increases, the insurer would reduce his/her investment in the stock and cede more liabilities to reinsurers at the same time, by considering the higher risk in both stocks and liabilities. This line of reasoning is also confirmed by the bottom graphs of Figure~\ref{fig:l pi to h} in the case where stock 1 has defaulted and the default intensity of stock 2 will increase (i.e., it corresponds to the default state $z=(1,0)$).
\begin{center}
\makeatletter
\def\@captype{figure}
\makeatother
\begin{figure}[htb!]
\hspace*{-10mm}
\begin{minipage}[t]{0.25\linewidth}
\centering
\includegraphics[width=3.3in]{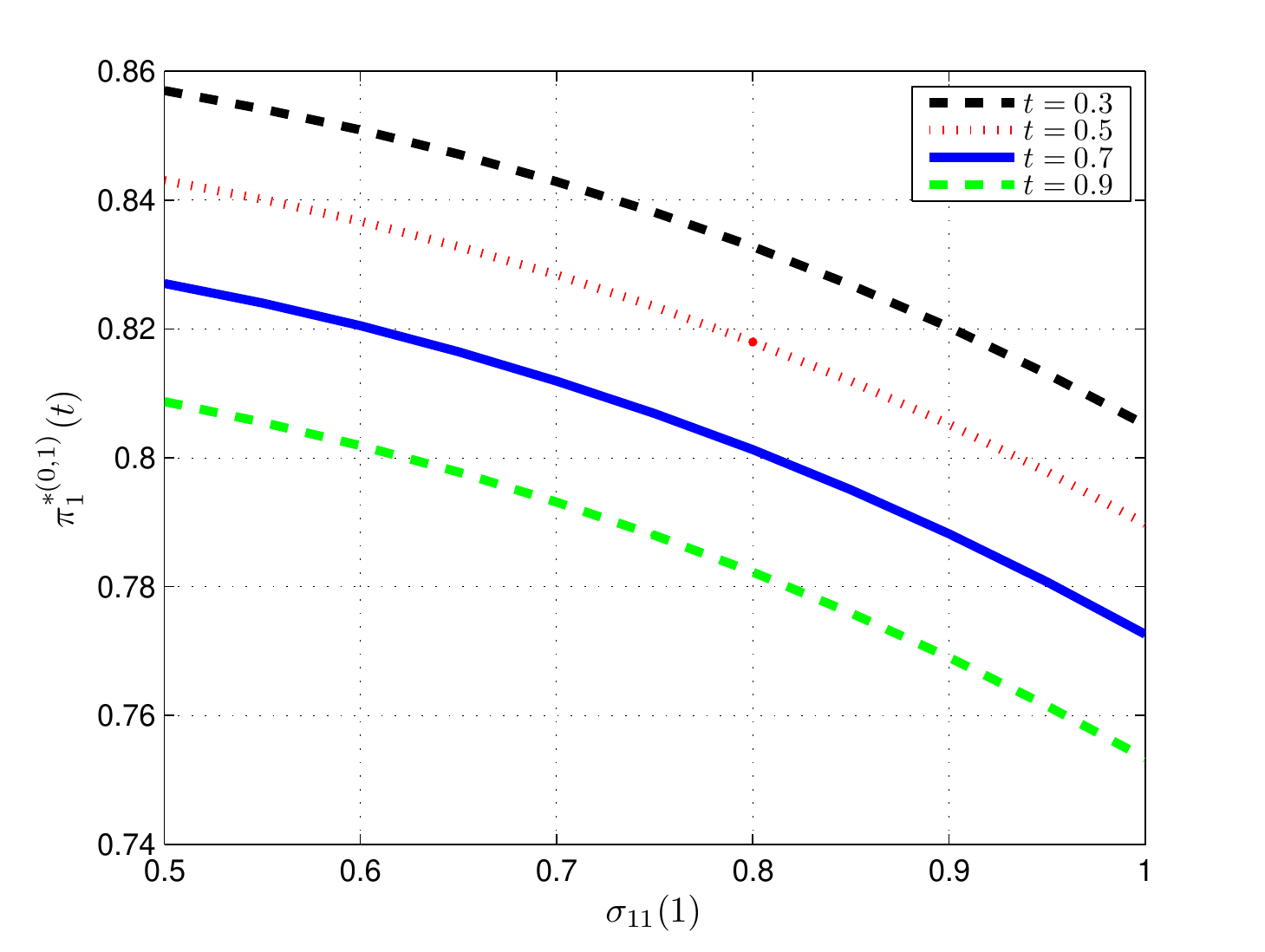}
\end{minipage}
\hspace{28ex}
\begin{minipage}[t]{0.25\linewidth}
\centering
\includegraphics[width=3.3in]{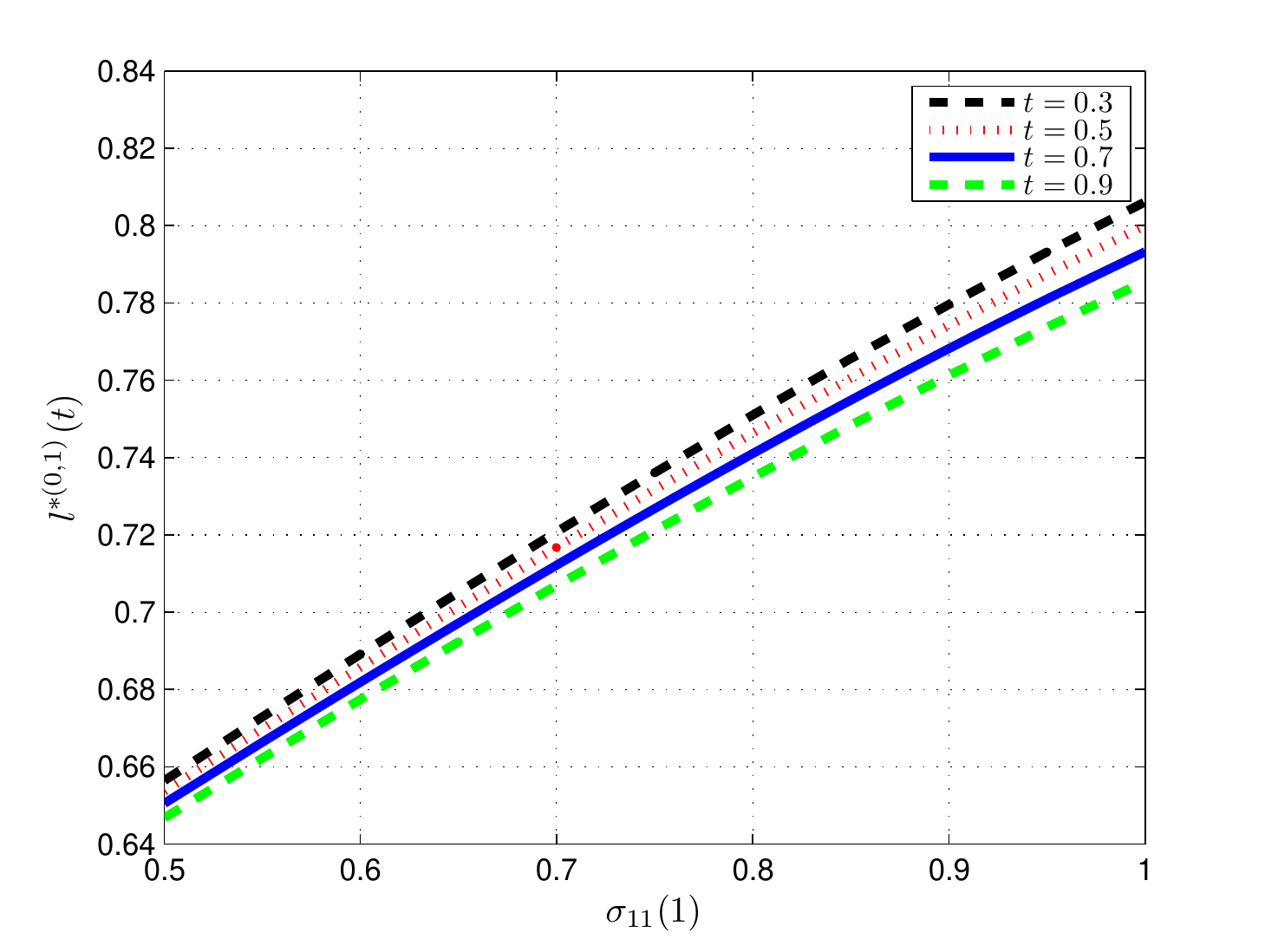}
\end{minipage}

\hspace*{-10mm}
\begin{minipage}[t]{0.25\linewidth}
\centering
\includegraphics[width=3.3in]{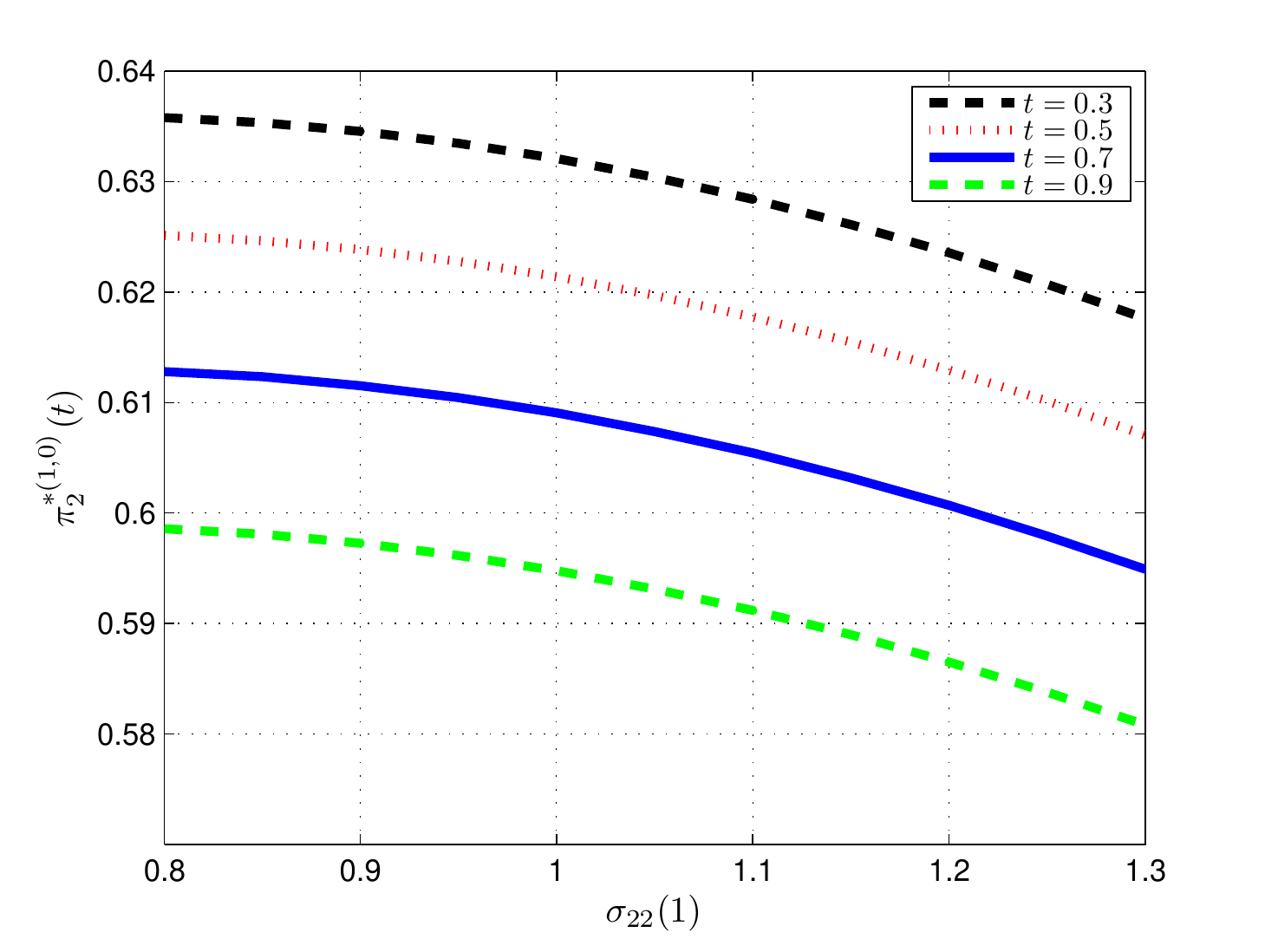}
\end{minipage}
\hspace{28ex}
\begin{minipage}[t]{0.25\linewidth}
\centering
\includegraphics[width=3.3in]{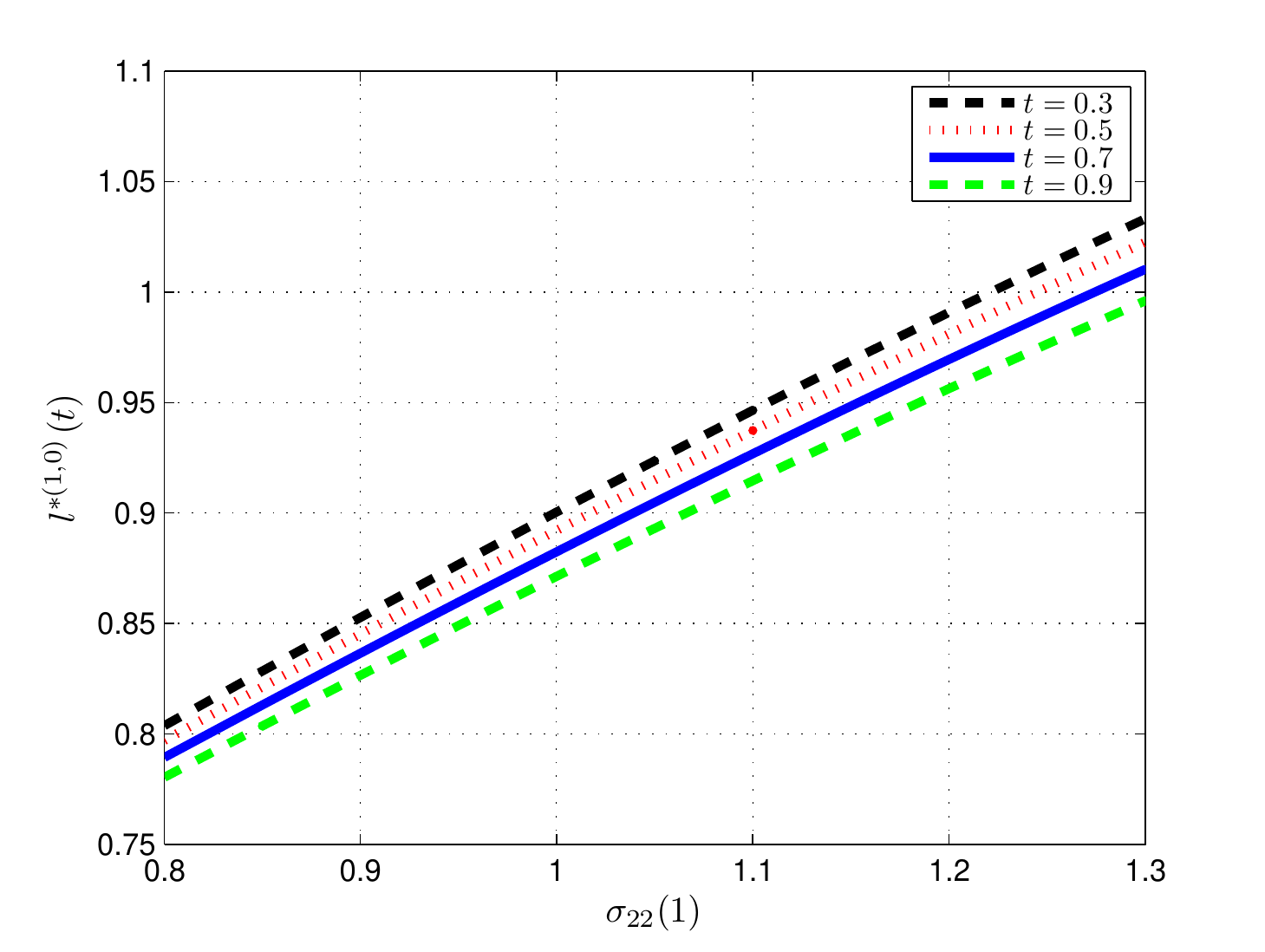}
\end{minipage}
\caption{\small Dependence of optimal strategies of stocks and risk control in regime 1 on volatility of stocks at different times.}\label{optm_strtgy_to_sigma}
\end{figure}
\end{center}

We next give an illustration of how market volatility impacts the optimal investment strategy of stocks and risk control.
Figure \ref{optm_strtgy_to_sigma} plots the optimal strategy of stocks and risk control in regime 1 at different times when the volatility
of stocks varies. The default states considered here are $z=(0,1)$ and $(1,0)$. A comparison between the left panel and the right panel of Figure \ref{optm_strtgy_to_sigma}
shows that the insurer decreases his/her investment in stocks and allocates a larger proportion of wealth to the liability, when the volatility of stocks increases. This happens because
a higher volatility induces the insurer to reduce his/her investment in the defaultable stocks and increase the proportion of wealth allocated to the liability. This can be also confirmed from the right panel of Figure \ref{optm_strtgy_to_sigma}. It also demonstrates that the optimal strategy for the liability is more sensitive to the changes of volatility of stocks than that to the changes in time. Consequently, the above comparison exploits that the optimal strategy of the liability is more sensitive to the changes in risk than that to the changes in time.
\begin{center}
\makeatletter
\def\@captype{figure}
\makeatother
\begin{figure}[htb!]
\hspace*{-10mm}
\begin{minipage}[t]{0.25\linewidth}
\centering
\includegraphics[width=3.5in]{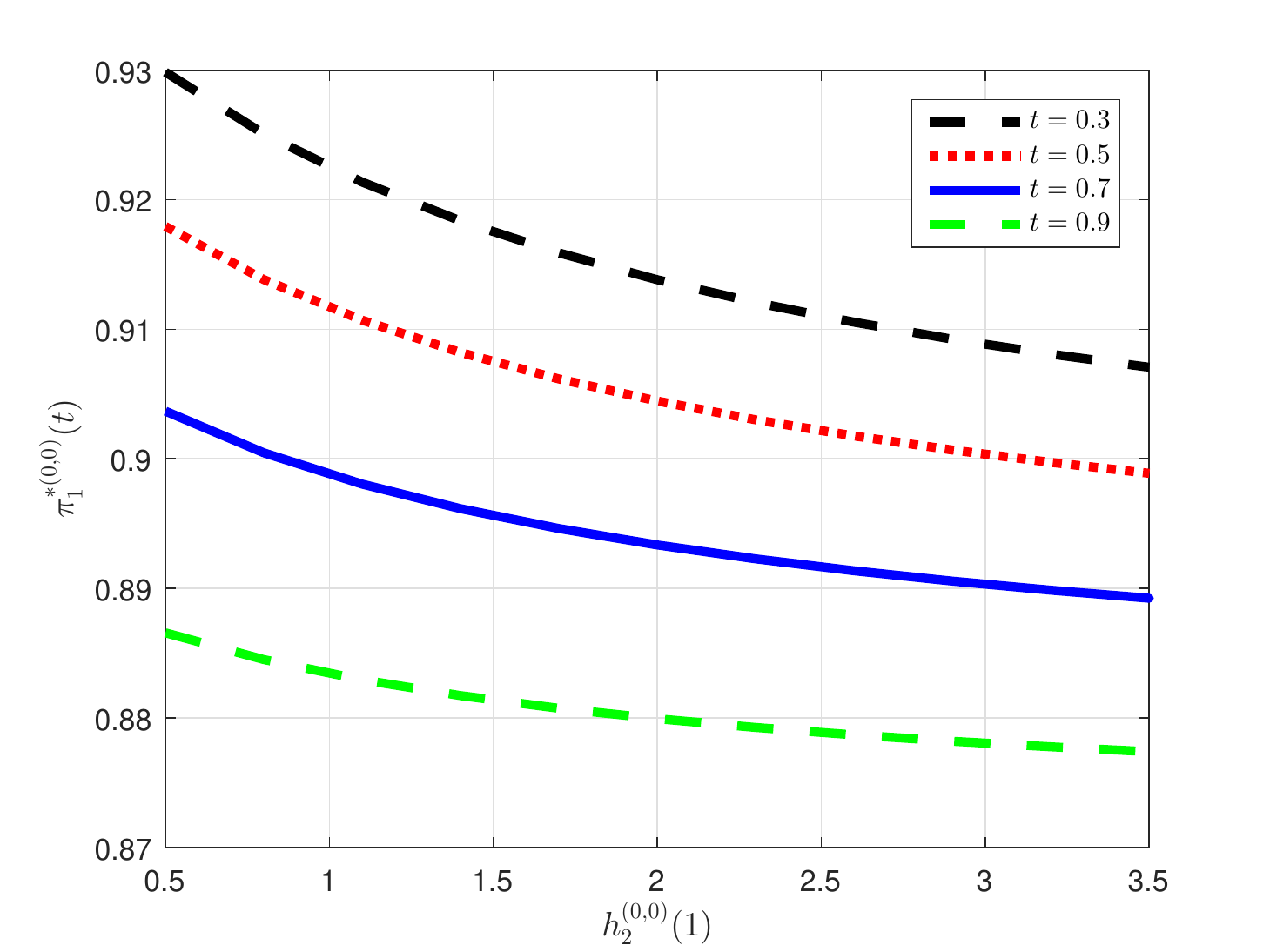}
\end{minipage}
\hspace{28ex}
\begin{minipage}[t]{0.25\linewidth}
\centering
\includegraphics[width=3.5in]{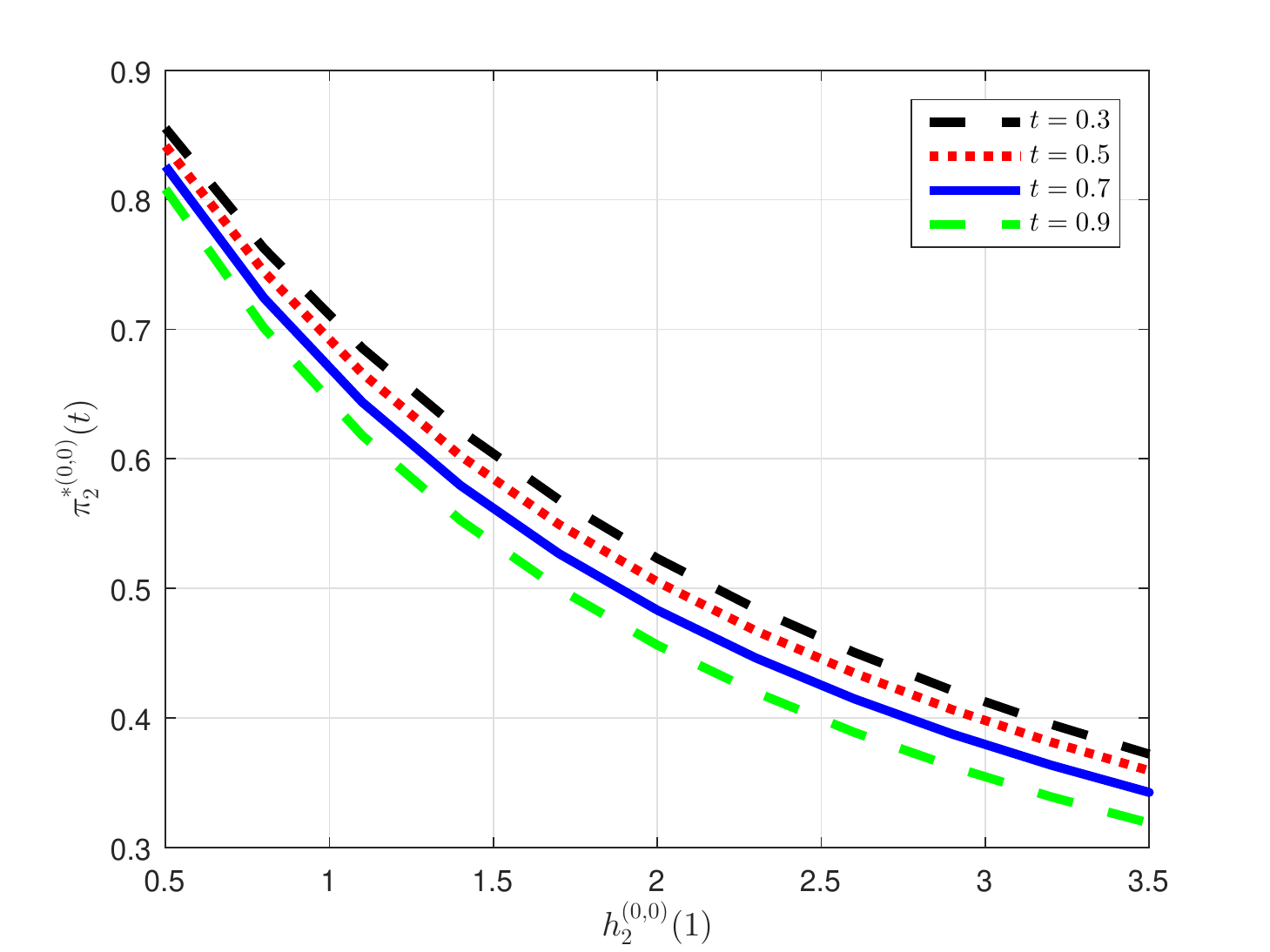}
\end{minipage}
\caption{\small Dependence of the optimal strategy of both stocks on the default intensity $h^{(0,0)}_2(1)$ of stock $2$ in regime 1. The current default state $z=(0,0)$, i.e. both stocks are alive. Left panel: dependence of the optimal strategy of stock $1$ on the default intensity $h^{(0,0)}_2(1)$ of stock $2$ in regime 1; Right panel: dependence of the optimal strategy of stock $2$ on the default intensity $h^{(0,0)}_2(1)$ of stock $2$ in regime 1.}\label{pi_ratio_to_h2}
\end{figure}
\end{center}

\begin{center}
\makeatletter
\def\@captype{figure}
\makeatother
\begin{figure}[htb!]
\hspace*{-10mm}
\begin{minipage}[t]{0.25\linewidth}
\centering
\includegraphics[width=3.5in]{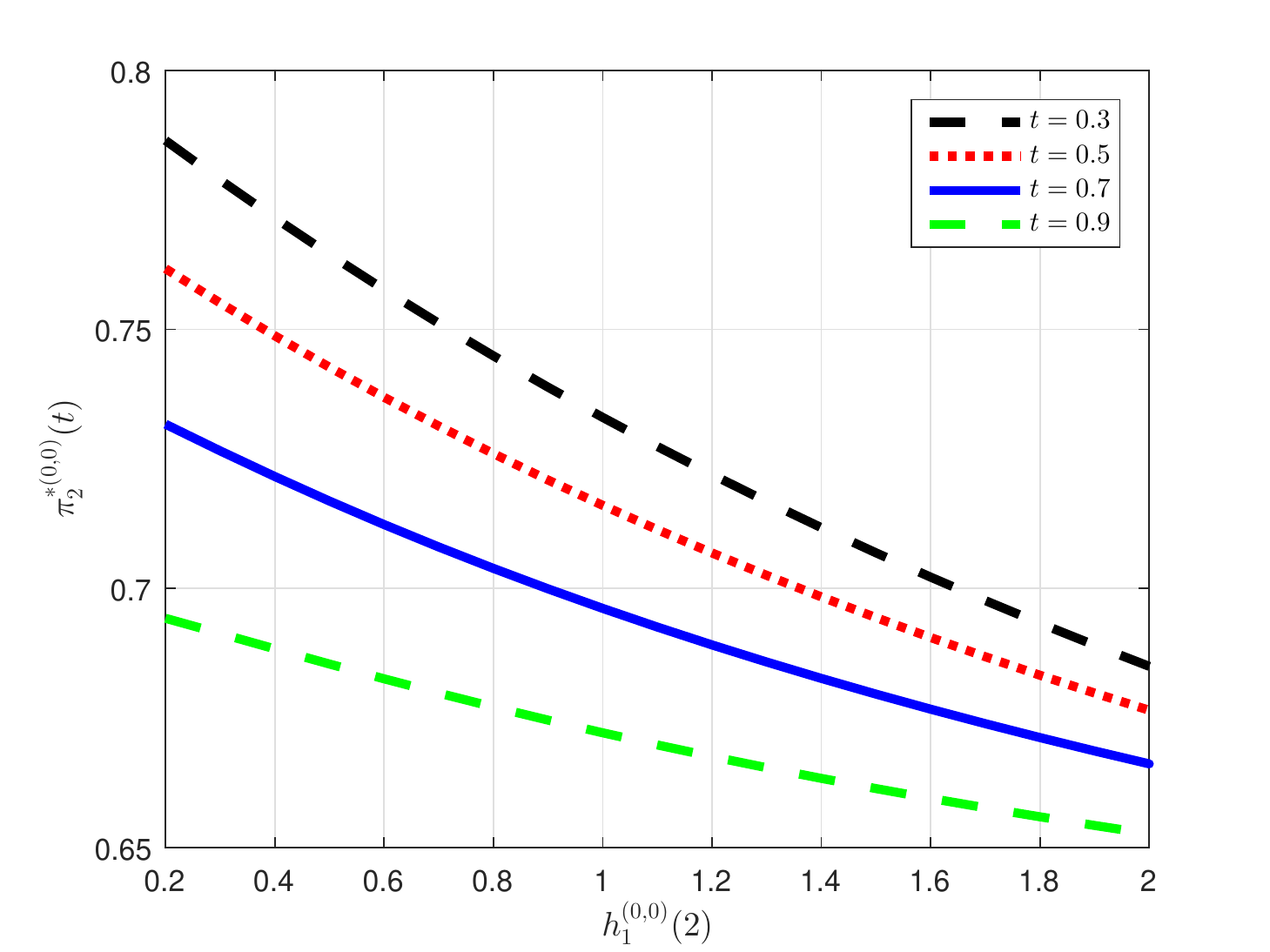}
\end{minipage}
\hspace{28ex}
\begin{minipage}[t]{0.25\linewidth}
\centering
\includegraphics[width=3.5in]{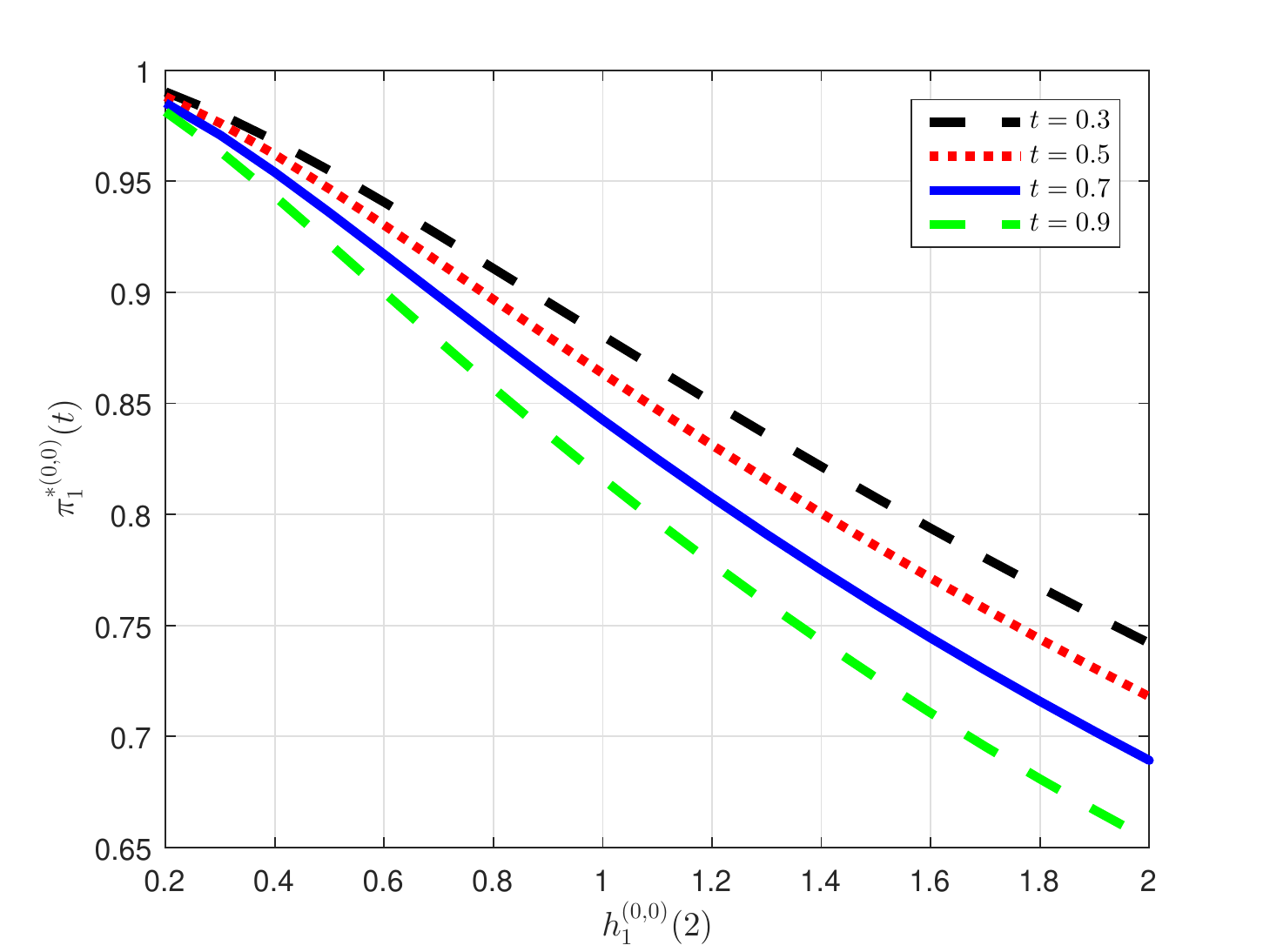}
\end{minipage}
\caption{\small Dependence of the optimal strategy of both stocks on the default intensity $h^{(0,0)}_1(2)$ of stock $1$ in regime 2. The current default state $z=(0,0)$, i.e. both stocks are alive. Left panel: dependence of the optimal strategy of stock $2$ in regime $2$ on the default intensity $h^{(0,0)}_1(2)$ of stock $1$ in regime 2; Right panel: dependence of the optimal strategy of stock $1$ in regime $2$ on the default intensity $h^{(0,0)}_1(2)$ of stock $1$ in regime 2.}\label{pi_ratio_to_h}
\end{figure}
\end{center}

\begin{center}
\makeatletter
\def\@captype{figure}
\makeatother
\begin{figure}[htb!]
\hspace*{-10mm}
\begin{minipage}[t]{0.25\linewidth}
\centering
\includegraphics[width=3.5in]{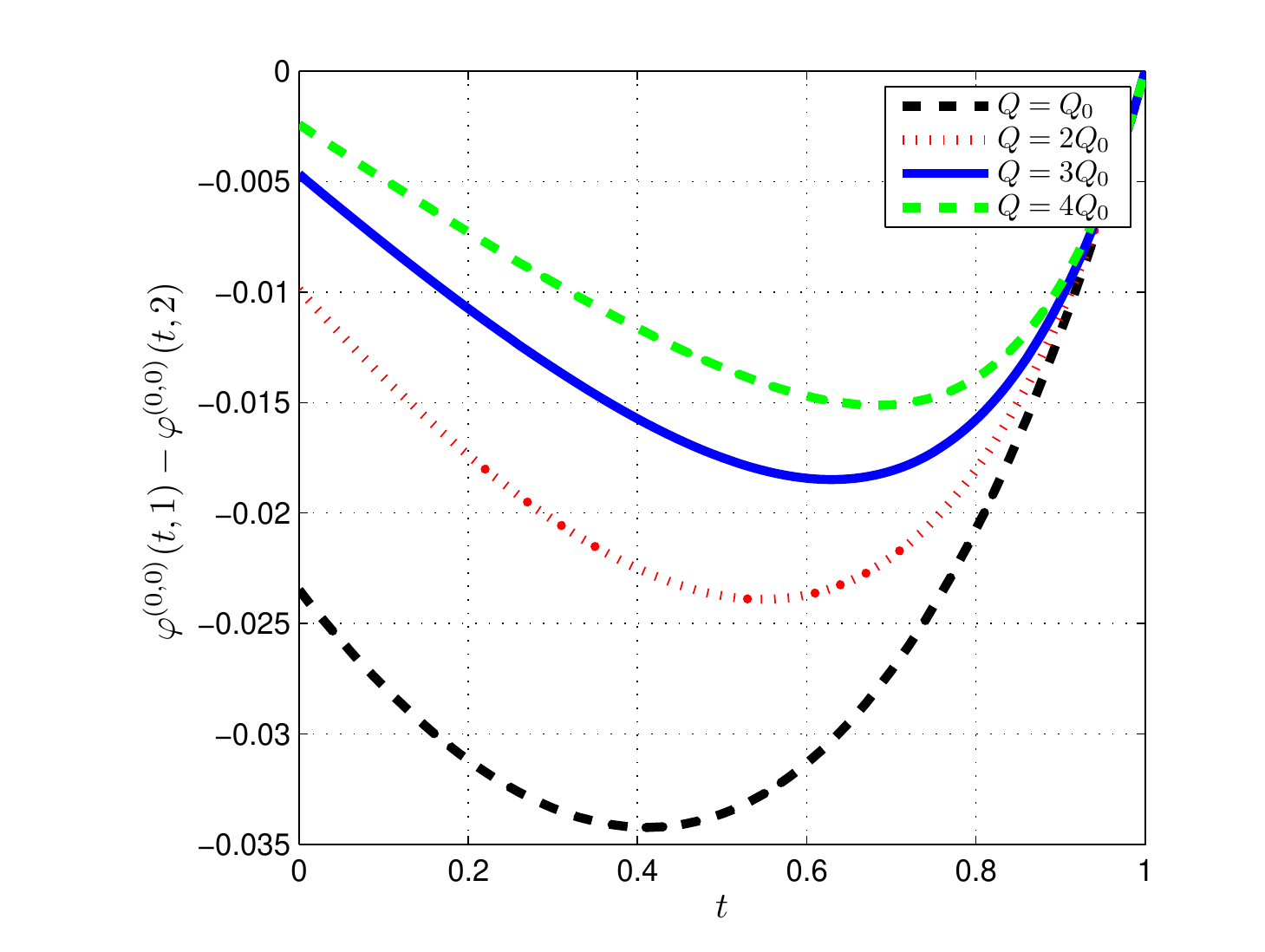}
\end{minipage}
\hspace{28ex}
\begin{minipage}[t]{0.25\linewidth}
\centering
\includegraphics[width=3.5in]{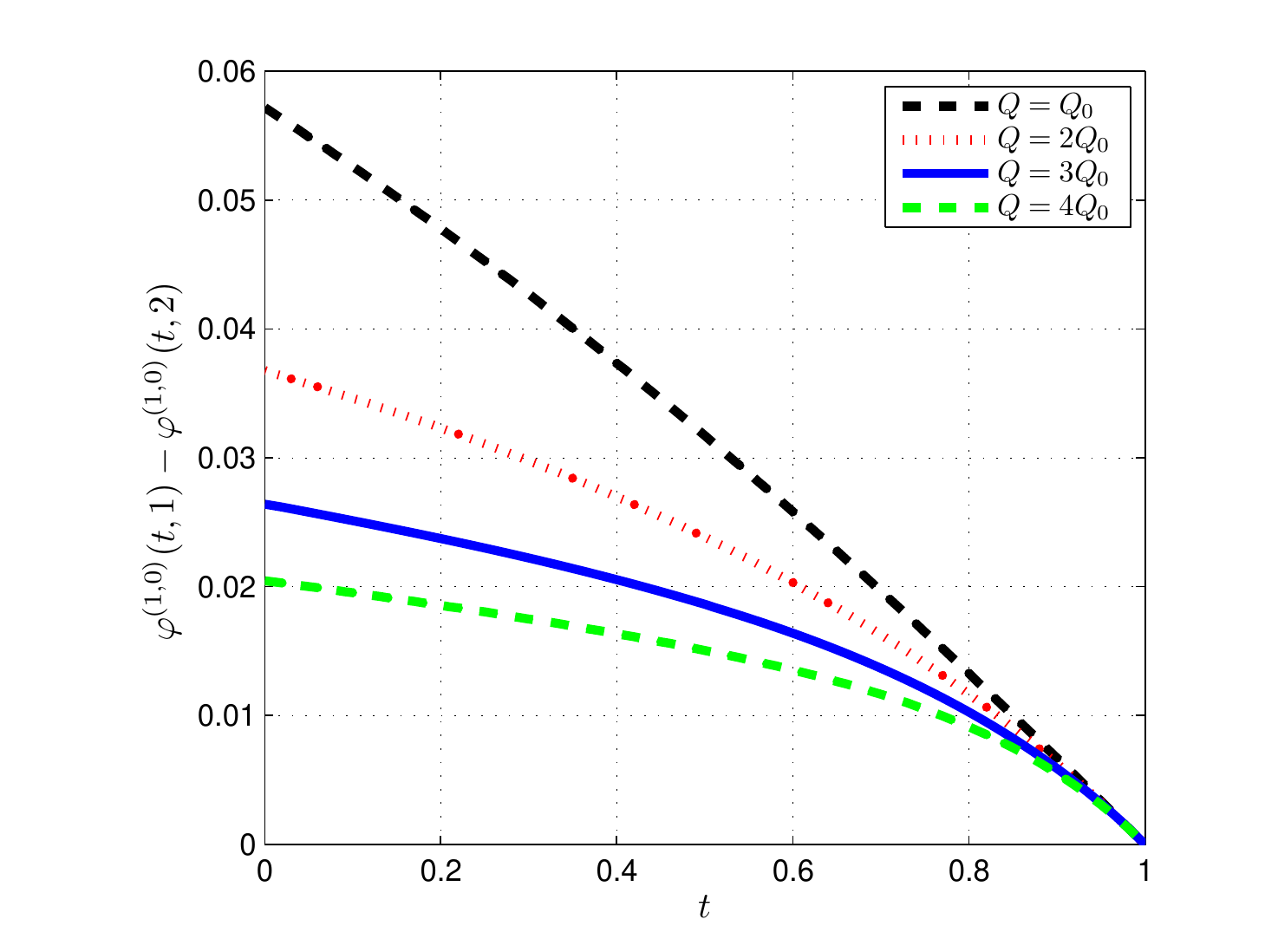}
\end{minipage}

\hspace*{-10mm}
\begin{minipage}[t]{0.25\linewidth}
\centering
\includegraphics[width=3.5in]{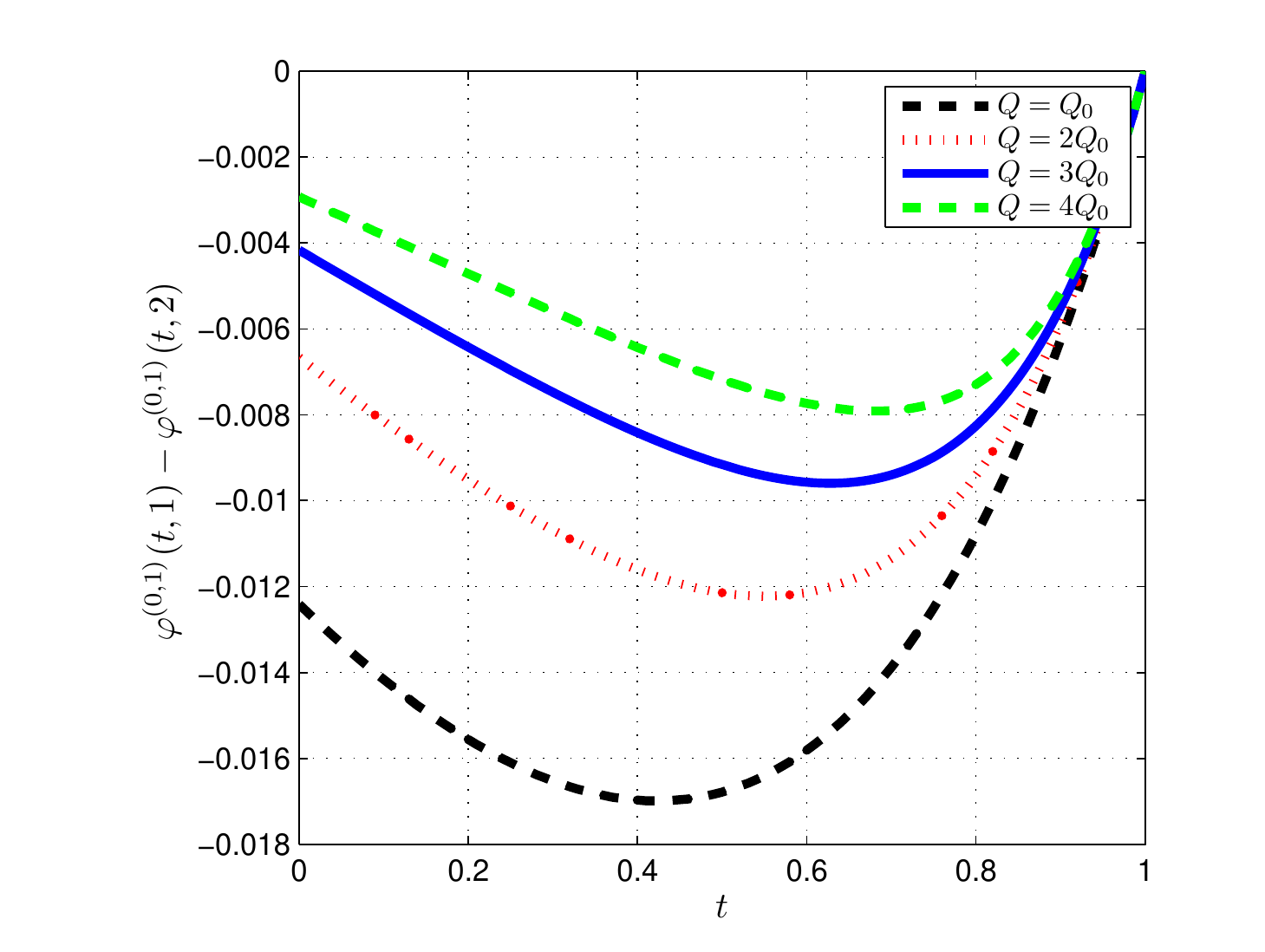}
\end{minipage}
\hspace{28ex}
\begin{minipage}[t]{0.25\linewidth}
\centering
\includegraphics[width=3.5in]{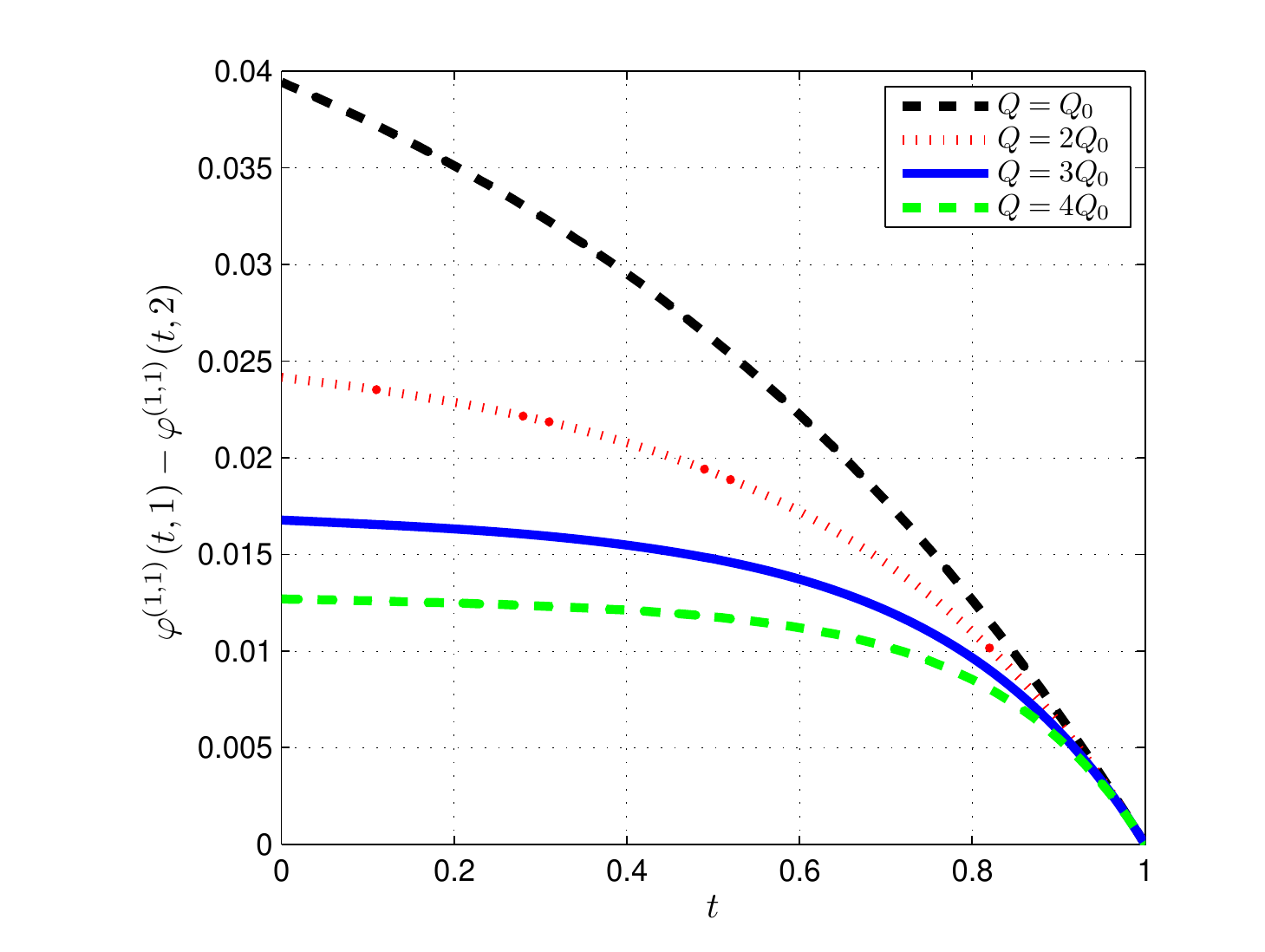}
\end{minipage}
\caption{\small The difference of value functions between two regimes given different generators of Markov chain $Y$ at different default states $z=(0,0)$, $(1,0)$, $(0,1)$ and $(0,0)$.}\label{varphi_differ}
\end{figure}
\end{center}

We finally assess the impact of default contagion on the optimal investment strategy of stocks and the value function respectively.
In particular, we explain how to disentangle the direct and indirect (contagion) effects of an increase in the default intensity.
Figure \ref{pi_ratio_to_h2} and \ref{pi_ratio_to_h} illustrate how default contagion impacts the investment strategy of the stock.
They suggest that when the default intensity of one stock increases, the insurer tends to reduce his/her investment in both stocks when both stocks are alive. This fact reflects the contagion property of default in this model: when one asset has higher default probability, the contagion property of default makes the investor reduce his/her investment in the other asset as well.
As it appears from the left panel of Figure~\ref{pi_ratio_to_h2}, when the default contagion of stock 2 increases, the insurer decreases the proportion of wealth allocated to stock 1.
This occurs because at the default of stock 2, the default intensity of stock 1 will instantaneously increase (an upward jump in the default intensity from $h_1^{(0,0)}=(0.5,0.75)$ to $h_1^{(0,1)}=(0.7,1)$), inducing a higher default risk of stock 1. Consequently, the risk averse insurer would allocate a smaller proportion of wealth to this stock.
Notice that at the default of stock 1, the default intensity of stock 2 will instantaneously increase because there is an upward jump in the default intensity from $h_2^{(0,0)}=(0.75,1.1)$ to $h_2^{(1,0)}=(0.9,1.3)$. The right panel of Figure~\ref{pi_ratio_to_h} confirms a similar trend for stock 2, however, the indirect contagion effect becomes more pronounced for the case of stock 2.

The direct effect of the default intensity is shown in the left panel of Figure \ref{pi_ratio_to_h2} (resp. in the right panel of Figure \ref{pi_ratio_to_h}). For a fixed default intensity of stock 1 (resp. stock 2), the insurer will invest less wealth in stock 1 (resp. stock 2) when the time to maturity decreases. In this regard,  it might be noted that the conditional survival probability of stocks $\Px(\tau_i>T|{\cal G}_t)$ with $t<T$ is given by ${\bf1}_{\tau_i>t}\Ex[e^{-\int_t^Th_i^{Z(s)}(Y(s))ds}|\G_t]$. As expected, this probability is decreasing with respect to the default intensity $h_i^z$ and for shorter time to maturity, all else being equal, the rate of change of this probability with respect to the default intensity becomes smaller (i.e., all else being equal, the conditional survival probability is not too sensitive to the default intensity when $t$ tends to $T$). Therefore, at an increase in the default intensity of stock 1 (resp. stock 2), the insurer tends to decrease investment in stock $1$ (resp. stock $2$) for shorter time to maturity. Moreover, the insurer will allocate less proportion of his/her wealth to stock 1 (resp. stock 2) as the default intensity of stock 1 (resp. stock 2) increases. Similar observation has also been made in Jiao, et al.~\cite{Jiao2013}.

Figure \ref{varphi_differ} depicts the difference of value functions between two regimes at four different default states $z=(0,0)$, $(1,0)$, $(0,1)$ and $(1,1)$ respectively. The graphs in Figure \ref{varphi_differ} confirm how the change of the absolute values of elements in the generator $Q$ affects the difference of value functions between two regimes.
At each default state, the difference of value functions between two regimes becomes tinier for larger absolute values of elements in the generator $Q$. This happens because a larger absolute value of the elements in $Q$ will result in a more frequent regime switching of the Markov chain. Consequently, the insurer relies more on his/her investment strategy
rather than the regime he/she is in when faced with a market with frequent regime switching.

\section*{Acknowledgments}
The authors gratefully acknowledge the constructive and insightful comments provided by one anonymous reviewer and Editor-in-Chief, Prof. Ulrich Horst, which helped to greatly improve the quality of the manuscript. This research of L. Bo and H. Liao
was supported in part by the NSF of China under Grant 11471254, The Key Research Program of Frontier
Sciences, CAS under Grant QYZDB-SSW-SYS009, and Fundamental Research Funds for Central
Universities under Grant WK3470000008.

\appendix
\section{Technical Proofs}\label{app:proof1}
\renewcommand\theequation{A.\arabic{equation}}
\setcounter{equation}{0}

\noindent{\it Proof of Lemma~\ref{lem:sol-hjben2}.}\quad Define $f(x)=Bx$ for $x\in\R^m$. By virtue of Proposition 1.1 of Charter 3 in Smith~\cite{smith08}, it suffices to verify that $f:\R^m\to\R^m$ is of type $K$, i.e., for any $x,y\in\R^m$ satisfying $x\leq y$ and $x_i=y_i$ for some $i=1,\ldots,m$, then $f_i(x)\leq f_i(y)$. Notice that $b_{ij}\geq0$ for all $i\neq j$. Then, it holds that
\begin{align}\label{eq:111}
f_i(x)&=(Bx)_i=\sum_{j=1}^mb_{ij}x_j=b_{ii}x_i+\sum_{j=1,j\neq i}^mb_{ij}x_j\nonumber\\
&=b_{ii}y_i+\sum_{j=1,j\neq i}^mb_{ij}x_j
\leq b_{ii}y_i+\sum_{j=1,j\neq i}^mb_{ij}y_j=f_i(y),
\end{align}
and hence $f$ is of type $K$. Thus, we complete the proof of the lemma. \hfill$\Box$\\

\noindent{\it Proof of Lemma~\ref{lem:sol-hjben}.}\quad The expression of the solution $\varphi(t,e_n)$ given by \eqref{eq:varphien} is obvious. Notice that $e_m\gg0$ and $q_{ij}\geq0$ for all $i\neq j$ since $Q=(q_{ij})_{m\times m}$ is the generator of the Markov chain. Then, in order to prove $\varphi(t,e_n)\gg0$ for all $t\in[0,T]$, using Lemma~\ref{lem:sol-hjben2}, it suffices to verify $[A^{(n)}]_{ij}\geq0$ for all $i\neq j$, however, $[A^{(n)}]_{ij}=q_{ij}$ for all $i\neq j$ using \eqref{eq:Aen}. Thus, we have verified the condition given in Lemma~\ref{lem:sol-hjben2}, and hence $\varphi(t,e_n)\gg0$ for all $t\in[0,T]$. \hfill$\Box$\\

\noindent{\it Proof of Lemma~\ref{lem:Gkesti}.}\quad It suffices to prove that, for any $x,y\in\R^m$ satisfying $x,y\geq\varepsilon e_m^{\top}$ with $\varepsilon>0$, there exists a constant $C=C(\varepsilon)>0$ depending on $\varepsilon>0$ only such that $|G_i^{(k)}(t,x)-G_i^{(k)}(t,y)|\leq C\| x-y\|$ for each $i=1,\ldots,m$. Since $\sigma(i)\sigma(i)^{\top}$ is also positive definite, $\sigma^{(k)}(i)\sigma^{(k)}(i)^{\top}$ is positive definite. Hence, there exists a constant $\delta>0$ such that $(\pi^{(k)})^{\top}\sigma^{(k)}(i)\sigma^{(k)}(i)^{\top}\pi^{(k)}\geq \delta\|\pi^{(k)}\|^2$. Then, for any $(\pi^{(k)},l)\in{\cal U}^{(k)}$, there exists a positive constant $C_1>0$ such that
\begin{align}\label{eq:est1}
&\frac{\gamma(\gamma-1)}2\left\{(\pi^{(k)})^{\top}\sigma^{(k)}(i)\sigma^{(k)}(i)^{\top}\pi^{(k)}+l^2\big(\phi(i)\phi(i)^{\top}+\bar{\phi}(i)\bar{\phi}(i)^{\top}\big)-2l(\pi^{(k)})^{\top}\sigma^{(k)}(i)\phi(i)\right\}\nonumber\\
&\qquad=\frac{\gamma(\gamma-1)}2\left\{l^2\bar{\phi}(i)\bar{\phi}(i)^{\top}+\|\sigma^{(k)}(i)^\top\pi^{(k)}-l\phi(i)\|^2\right\}\nonumber\\
&\qquad\leq\frac{\gamma(\gamma-1)}2\left\{\alpha l^2\bar{\phi}(i)\bar{\phi}(i)^{\top}+\frac12(1-\alpha)\|\sigma^{(k)}(i)^\top\pi^{(k)}\|^2-(1-\alpha)\|l\phi(i)\|^2\right\}\nonumber\\
&\qquad=\frac{\gamma(\gamma-1)}2\left\{l^2(\alpha\bar{\phi}(i)\bar{\phi}(i)^{\top}-(1-\alpha)\|\phi(i)\|^2)+\frac12(1-\alpha)\|\sigma^{(k)}(i)^\top\pi^{(k)}\|^2\right\}\nonumber\\
&\qquad\leq-C_1(\|\pi^{(k)}\|^2+l^2),
\end{align}
where the constant $\alpha\in(\max_{i=1,\ldots,m}\big\{\frac{\|\phi(i)\|^2}{\bar{\phi}(i)\bar{\phi}(i)^{\top}+\|\phi(i)\|^2}\big\},1)$. 
On the other hand, for any $(\pi^{(k)},l)\in{\cal U}^{(k)}$, it holds that
\begin{align}\label{eq:est2}
\gamma\big\{(\pi^{(k)})^{\top}\theta^{(k)}(i)+(p^{(k)}(i)-c(i))l\big\}&\leq\gamma\|\theta^{(k)}(i)\|\|\pi^{(k)}\|+\gamma|p^{(k)}(i)-c(i)|l\nonumber\\
&\leq C_2\sqrt{\|\pi^{(k)}\|^2+l^2},
\end{align}
where the constant $C_2:=\max_{i=1,\ldots,m}\big\{\gamma\sqrt{\|\theta^{(k)}(i)\|^2+|p^{(k)}(i)-c(i)|^2}\big\}>0$. Finally, for any $(\pi^{(k)},l)\in{\cal U}^{(k)}$, we have that $\{(1-lg(i))^\gamma-1\}\nu^{(k)}(i)\leq C_3l$, where the constant $C_3:=\max_{i=1,\ldots,m}\{\gamma g(i)\nu^{(k)}(i)\}>0$. Then, by virtue of \eqref{eq:Hk}, it follows that, for any $(\pi^{(k)},l)\in{\cal U}^{(k)}$ and $i=1,\ldots,m$,
\begin{align}\label{eq:esti4}
H^{(k)}((\pi^{(k)},l),i)\leq -C_1\big(\|\pi^{(k)}\|^2+l^2\big)+C_4\sqrt{\|\pi^{(k)}\|^2+l^2}.
\end{align}
Here $C_4=C_2+C_3$. This yields that there exists a constant $C_5>0$ such that when $(\pi^{(k)},l)\in{\cal U}^{(k)}$ and $\|\pi^{(k)}\|^2+l^2>C_5$, we have $H^{(k)}((\pi^{(k)},l),i)<0$ for all $i=1,\ldots,m$, and meanwhile, for $x\geq \varepsilon e_{m}^{\top}$ with $\varepsilon>0$,
\begin{align}\label{eq:est5}
&\sum_{j\notin\{j_1,\ldots,j_k\}}(1-\pi_j^{(k)})^\gamma h_{j}^{(k)}(i)\varphi^{(l+1),j}(t,i)+H^{(k)}((\pi^{(k)},l),i)x_i\nonumber\\
&\qquad\leq\big(1+\|\pi^{(k)}\|\big)^\gamma \sum_{j\notin\{j_1,\ldots,j_k\}}h_{j}^{(k)}(i)\varphi^{(l+1),j}(t,i)+\varepsilon H^{(k)}((\pi^{(k)},l),i)\nonumber\\
&\qquad\leq C_6\big(1+\|\pi^{(k)}\|^\gamma\big)\sum_{j\notin\{j_1,\ldots,j_k\}}h^{(k)}_{j}(i)+\varepsilon\left\{-C_1(\|\pi^{(k)}\|^2+l^2)+C_4\sqrt{\|\pi^{(k)}\|^2+l^2}\right\}\nonumber\\
&\qquad\leq-\varepsilon C_1\big(\|\pi^{(k)}\|^2+l^2\big)+\varepsilon C_4\sqrt{\|\pi^{(k)}\|^2+l^2}+C_7\big(\sqrt{\|\pi^{(k)}\|^2+l^2}\big)^{\gamma}+C_8,
\end{align}
for some constants $C_6,C_7,C_8>0$. Notice that we used the recursive assumption that the HJB system \eqref{eq:hjbeqn} admits a positive unique (classical) solution $\varphi^{(k+1),j}(t)$ on $t\in[0,T]$ for $j\notin\{j_1,\ldots,j_k\}$. Then $\varphi^{(k+1),j}(t)$ is continuous on $[0,T]$, and hence $\varphi^{(k+1),j}(t)$ is bounded on $[0,T]$. From the estimate \eqref{eq:est5}, it follows that, for any $x\geq\varepsilon e_m^{\top}$, there exists a positive constant $C_9=C_9(\varepsilon)$ such that when $(\pi^{(k)},l)\in{\cal U}^{(k)}$, $\|\pi^{(k)}\|^2+l^2>C_9$ and $x\geq\varepsilon e_m^{\top}$, it holds that, for each $i=1,\ldots,m$,
\begin{align}\label{eq:est6}
&\sum_{j\notin\{j_1,\ldots,j_k\}}(1-\pi_j^{(k)})^\gamma h^{(k)}_{j}(i)\varphi^{(l+1),j}(t,i)+H^{(k)}((\pi^{(k)},l),i)x_i<0.
\end{align}
On the other hand, for $i=1,\ldots,m$, it holds that
\begin{align}\label{eq:G>0}
G^{(k)}_i(t,x)=&\sup_{(\pi^{(k)},l)\in{\cal U}^{(k)}}\left\{\sum_{j\notin\{j_1,\ldots,j_k\}}(1-\pi_{j}^{(k)})^\gamma h_{j}^{(k)}(i)\varphi^{(k+1),j}(t,i)+H^{(k)}((\pi^{(k)},l),i)x_i\right\}\nonumber\\
\geq&\sum_{j\notin\{j_1,\ldots,j_k\}} h^{(k)}_{j}(i)\varphi^{(k+1),j}(t,i)+H^{(k)}((0e_{n-k}^{\top},0),i)x_i\nonumber\\
=&\sum_{j\notin\{j_1,\ldots,j_k\}} h_{j}^{(k)}(i)\varphi^{(k+1),j}(t,i)>0.
\end{align}
Thus, using the estimate \eqref{eq:est6}, we have that, for all $x\geq\varepsilon e_m^{\top}$,
\begin{align}\label{eq:Gk2}
G^{(k)}_i(t,x)=&\sup_{\substack{(\pi^{(k)},l)\in{\cal U}^{(k)}\\\|\pi^{(k)}\|^2+l^2\leq C_9(\varepsilon)}}\left\{\sum_{j\notin\{j_1,\ldots,j_k\}}(1-\pi_{j}^{(k)})^\gamma h_{j}^{(k)}(i)\varphi^{(k+1),j}(t,i)+H^{(k)}((\pi^{(k)},l),i)x_i\right\}.
\end{align}
It follows from \eqref{eq:Gk2} and \eqref{eq:Hk} that, for all $x,y\geq\varepsilon e_m^{\top}$,
\begin{align}
G^{(k)}_i(t,x)=&\sup_{\substack{(\pi^{(k)},l)\in{\cal U}^{(k)}\\\|\pi^{(k)}\|^2+l^2\leq C_9(\varepsilon)}}\Bigg\{\sum_{j\notin\{j_1,\ldots,j_k\}}(1-\pi_{j}^{(k)})^\gamma h_{j}^{(k)}(i)\varphi^{(k+1),j}(t,i)+H^{(k)}((\pi^{(k)},l),i)y_i\nonumber\\
&\qquad\qquad\qquad+H^{(k)}((\pi^{(k)},l),i)(x_i-y_i)\Bigg\}\nonumber\\
\leq&\sup_{\substack{(\pi^{(k)},l)\in{\cal U}^{(k)}\\\|\pi^{(k)}\|^2+l^2\leq C_9(\varepsilon)}}\Bigg\{\sum_{j\notin\{j_1,\ldots,j_k\}}(1-\pi_{j}^{(k)})^\gamma h_{j}^{(k)}(i)\varphi^{(k+1),j}(t,i)+H^{(k)}((\pi^{(k)},l),i)y_i\nonumber\\
&\qquad\qquad\qquad+\left|H^{(k)}((\pi^{(k)},l),i)\right|\left|x_i-y_i\right|\Bigg\}\nonumber\\
\leq&G^{(k)}_i(t,y)+\left|x_i-y_i\right|\sup_{\substack{(\pi^{(k)},l)\in{\cal U}^{(k)}\\\|\pi^{(k)}\|^2+l^2\leq C_9(\varepsilon)}}\left\{\left|H^{(k)}((\pi^{(k)},l),i)\right|\right\}\nonumber\\
\leq&G^{(k)}_i(t,y)+C(\varepsilon)\left|x_i-y_i\right|,
\end{align}
where $C(\varepsilon)>0$ is a constant which depends on $\varepsilon>0$ only. Then, the above estimate results in the validity of the estimate \eqref{eq:Gkesti} for all $x,y\in\R^m$ satisfying $x,y\geq\varepsilon e_m^{\top}$. Thus, we complete the proof of the lemma. \hfill$\Box$\\

\noindent{\it Proof of Lemma~\ref{lem:comparison}.}\quad For $p>0$, let $g_{1}^{(p)}(t)=(g_{1i}^{(p)}(t);\ i=1,\ldots,m)^{\top}$ be the solution to the following dynamical system given by
\begin{equation}
\left\{
\begin{aligned}
\frac d{dt}g_{1}^{(p)}(t)=&f(t,g_{1}^{(p)}(t))+\tilde{f}(t,g^{(p)}_{1}(t))+\frac{1}{p}e_m^{\top},\ \text{ in }(0,T];\\
g_{1}^{(p)}(0)=&\xi_1+\frac{1}{p}e_m^{\top}.
\end{aligned}
\right.
\end{equation}
Then, for all $t\in(0,T]$, it holds that
\begin{align*}
\|g_{1}^{(p)}(t)-g_1(t)\|\leq&\|g_{1}^{(p)}(0)-g_1(0)\|+\int_0^t\big\|f(s,g_{1}^{(p)}(s))-f(s,g_1(s))\big\|ds\nonumber\\
&+\int_0^t\big\|\tilde{f}(s,g_{1}^{(p)}(s))-\tilde{f}(s,g_1(s))\big\|ds+\frac1p\int_0^t\|e_m\|ds\nonumber\\
\leq&\frac2p\|e_m\|+(C+\tilde{C})\int_0^t\big\|g_{1}^{(p)}(s)-g_1(s)\big\|ds.
\end{align*}
Here $C>0$ (resp. $\tilde{C}>0$) is the Lipschitz constant of $f(t,x)$ (resp. $\tilde{f}(t,x)$) in $x$. Then,
the Gronwall's lemma yields that $g_{1}^{(p)}(t)\to g_1(t)$ for all $t\in[0,T]$ as $p\to\infty$.
We claim that $g_{1}^{(p)}(t)\gg g_2(t)$ for all $t\in[0,T]$. If the claim were false, notice that $g_{1}^{(p)}(0)\gg g_2(0)$, and $g_1^{(p)}(t),g_2(t)$ are continuous on $[0,T]$, then there exists a $t_0\in(0,T]$ such that $g_{1}^{(p)}(s)\geq g_2(s)$ on $s\in[0,t_0]$ and $g_{1i}^{(p)}(t_0)=g_{2i}(t_0)$ for some $i\in\{1,\ldots,m\}$. Since $t_0>0$, $g_1^{(p)}(t)$ and $g_2(t)$ are differentiable on $(0,T]$, we have that
\begin{align*}
\frac d{dt}g_{1i}^{(p)}(t)\big|_{t=t_0}=\lim_{\epsilon\to0}\frac{g_{1i}^{(p)}(t_0)-g_{1i}^{(p)}(t_0-\epsilon)}{\epsilon}
\leq\lim_{\epsilon\to0}\frac{g_{2i}(t_0)-g_{2i}(t_0-\epsilon)}{\epsilon}= \frac d{dt}g_{2i}(t)\big|_{t=t_0}.
\end{align*}
On the other hand, since $f(t,\cdot)$ satisfies the type $K$ condition for each $t\in[0,T]$ and $\tilde{f}(t,x)\geq0$ for all $(t,x)\in[0,T]\times\R^m$, for the above $i$, we also have that
\begin{align}
\frac d{dt}g_{1i}^{(p)}(t)\big|_{t=t_0}=&f_i(t_0,g_{1i}^{(p)}(t_0))+\tilde{f}_i(t_0,g_{1}^{(p)}(t_0))+\frac1p\nonumber\\
>&f_i(t_0,g_{1i}^{(p)}(t_0))\geq f_i(t_0,g_2(t_0))=\frac d{dt}g_{2i}(t)\big|_{t=t_0}.
\end{align}
This results in a contradiction, and hence $g_{1}^{(p)}(t)\gg g_2(t)$ for all $t\in[0,T]$. Thus, it holds that $g_1(t)\geq g_2(t)$ for all $t\in[0,T]$ by letting $p$ tend to infinity. \hfill$\Box$\\

\noindent{\it Proof of Theorem~\ref{thm:vefi}.}\quad For $(t,x,i,z)\in[0,T]\times\R_+\times\{1,\ldots,m\}\times{\cal S}$, note that $\varphi(T,i,z)=\frac{1}{\gamma}$. Then, by virtue of It\^o's formula, for all $(\pi,l)\in\tilde{\cal U}$, it follows that
\begin{align*}
&\frac{1}{\gamma}(X^{\pi,l}(T))^\gamma=(X^{\pi,l}(t))^\gamma \varphi(t,Y(t),Z(t))+\int_t^T(X^{\pi,l}(s))^\gamma\frac{\partial \varphi(s,Y(s),Z(s))}{\partial s}ds\nonumber\\
&\quad+\int_t^T\gamma (X^{\pi,l}_s)^{\gamma-1}\varphi(s,Y(s),Z(s))dX^{\pi,l}(s)^c\\
&\quad+\frac{\gamma(\gamma-1)}{2}\int_t^T(X^{\pi,l}(s))^{\gamma-2}\varphi(s,Y(s),Z(s))d[X^{\pi,l},X^{\pi,l}]^c(s)\\
&\quad+\int_t^T\varphi(s,Y(s-),Z(s-))(X^{\pi,l}(s-))^\gamma[(1-l(s)g(Y(s-)))^\gamma-1]dN(s)\\
&\quad+\sum_{j=1}^n\int_t^T(X^{\pi,l}(s-))^\gamma\big[(1-\pi_j(s-))^\gamma  \varphi(s,Y(s-),Z^j(s-))-\varphi(s,Y(s-),Z(s-))\big]dZ_j(s)\\
&\quad+\int_t^T\sum_{j\neq Y(s-)}(X^{\pi,l}(s-))^\gamma \big[\varphi(s,j,Z(s-))-\varphi(s,Y(s-),Z(s-))\big]dH_{Y(s-),j}(s)\\
&\quad=(X^{\pi,l}(t))^\gamma \varphi(t,Y(t),Z(t))+\int_t^T(X^{\pi,l}(s))^\gamma{\cal A}(\pi,l;s,Y(s),Z(s))ds+M^{\pi,l}(T)-M^{\pi,l}(t).
\end{align*}
Here for $(\pi,l)\in(-\infty,1]^n\times[0,\infty)$ and $(t,i,z)\in[0,T]\times\{1,\ldots,m\}\times{\cal S}$, the coefficient is given by
\begin{align*}
&{\cal A}(\pi,l;t,i,z)\nonumber\\
&\quad=\frac{\partial\varphi(t,i,z)}{\partial t}+\Bigg\{\gamma\Big[r(i)+\pi^{\top}(I-diag(z))\theta(i,z)+\pi^{\top}(I-diag(z))h(i,z)+(p(i,z)-c(i))l\Big]\\
&\qquad+\frac{\gamma(\gamma-1)}2\Big[\pi^{\top}(I-diag(z))\sigma(i)\sigma(i)^{\top}(I-diag(z))\pi+l^2\big(\phi(i)\phi(i)^{\top}+\bar{\phi(i)}\bar{\phi(i)}^{\top}\big)\\
&\qquad-2l\pi^{\top}(I-diag(z))\sigma(i)\phi(i)^{\top}\Big]+[(1-lg(i))^\gamma-1]\nu(i,z)\bigg\}\varphi(t,i,z)\\
&\qquad+\sum_{j=1}^n[(1-\pi_j)^\gamma \varphi(t,i,z^j)-\varphi(t,i,z)](1-z_j)h_j(i,z)+\sum_{j\neq i}[\varphi(t,j,z)-\varphi(t,i,z)]q_{ij},
\end{align*}
and the $\Px$-(local) martingale is defined as
\begin{align*}
M^{\pi,l}(t)=&\int_0^t\gamma(X^{\pi,l}(s))^\gamma \varphi(s,Y(s),Z(s))\big[\pi(s)^{\top}(I-diag(Z(s)))\sigma(Y(s))-l(s)\phi(Y(s))\big]dW(s)\\
&+\int_0^t\gamma(X^{\pi,l}(s))^\gamma \varphi(s,Y(s),Z(s))l(s)\bar{\phi}(Y(s))d\bar{W}(s)\\
&+\int_0^t(X^{\pi,l}(s))^\gamma \varphi(s,Y(s-),Z(s-))[(1-l(s)g(Y(s-)))^\gamma-1]d\tilde{N}(s)\\
&+\sum_{j=1}^n\int_0^T(X^{\pi,l}(s-))^\gamma[(1-\pi_j(s))^\gamma \varphi(s,Y(s-),Z(s-)^j)-\varphi(s,Y(s-),Z(s-))]dM_j(s)\\
&+\int_0^t\sum_{j\neq Y(s-)}(X^{\pi,l}(s-))^\gamma \big[\varphi(s,j,Z(s-))-\varphi(s,Y(s-),Z(s-))]d\tilde{H}_{Y(s-),j}(s),
\end{align*}
where we used the following $\Px$-martingale processes given by, for $t\in[0,T]$,
\begin{align*}
\tilde{N}(t)&:=N(t)-\int_0^t\nu(Y(s),Z(s))ds,\quad
\tilde{H}_{ij}(t):=H_{ij}(t)-\int_0^tq_{ij}\mathds{1}_{Y(s)=i}ds,
\end{align*}
for all $i,j\in\{1,\ldots,m\}$ and $i\neq j$. Here, we recall that the process $H_{ij}(t)$ is defined by \eqref{eq:Hil}.
Using \eqref{eq:hjbeqn}, \eqref{eq:lnstar} and \eqref{eq:optimumk}, for $t\in[0,T]\times\{1,\ldots,m\}\times{\cal S}$, we have that ${\cal A}(\pi,l;t,i,z)\leq {\cal A}(\pi^*,l^*;t,i,z)=0$ for all $(\pi,l)\in{\cal U}$. Moreover, define $\tau_a:=\inf\{s\geq t;\ |X^{\pi,l}(s)|>a\}$ for $a>0$. Eq.~\eqref{eq:sdeX2} gives that, for $s\in[t,T]$,
\begin{align}\label{diff}
X^{\pi,l}(s\wedge\tau_a)=&X^{\pi,l}(s\wedge\tau_a-)\\
&\times[1-\tilde{\pi}^\top(s\wedge\tau_a)\Delta M(s\wedge\tau_a)-\tilde{l}(s\wedge\tau_a)g(Y(s\wedge\tau_a-))\Delta N(s\wedge\tau_a)],\nonumber
\end{align}
where the feedback controls are given by
\begin{align*}
\tilde{\pi}(s\wedge\tau_a)=&\pi\big(s\wedge\tau_a,X^{\pi,l}(s\wedge\tau_a-),Y(s\wedge\tau_a-),Z(s\wedge\tau_a-)\big),\\
\tilde{l}(s\wedge\tau_a)=&l\big(s\wedge\tau_a,X^{\pi,l}(s\wedge\tau_a-),Y(s\wedge\tau_a-),Z(s\wedge\tau_a-)\big).
\end{align*}
Notice that $(\pi,l)\in{\cal U}$ is locally bounded, and hence
\begin{align*}
|\tilde{\pi}(s\wedge\tau_a)|+|\tilde{l}(s\wedge\tau_a)|\leq C_1(\pi,l,a,T),\quad s\in[t,T].
\end{align*}
The positive constant $C_1$ depends on $(\pi,l)$, $a$ and $T$ only. Since $|\Delta M|\vee|\Delta N|\leq 1$, it follows that
\begin{align*}
|X^{\pi,l}(s\wedge\tau_a)|\leq C_2(\pi,l,a,T),\quad s\in[t,T],
\end{align*}
where $C_2$ is a positive constant which depends on $(\pi,l)$, $a$ and $T$ only. This implies that $M^{\pi,l}(\cdot\wedge\tau_a)$ is a $\Px$-martingale. Hence, it holds that
\begin{align}
\Ex_{t,x,i,z}\left[U(X^{\pi,l}(T\wedge\tau_a)\big)\right]&\leq x^\gamma \varphi(t,i,z)+\Ex_{t,x,i,z}\left[{M}^{\pi,l}(T\wedge\tau_a)-{M}^{\pi,l}(t)\right]\nonumber\\
&=x^\gamma \varphi(t,i,z),
\end{align}
where we set $\Ex_{t,i,z}[\cdot]:=\Ex[\cdot\mid X^{\pi,l}(t)=x,Y(t)=i,Z(t)=z]$ for $(t,x,i,z)\in[0,T]\times\R_+\times\{1,\ldots,m\}\times{\cal S}$. It follows from Fatou's lemma that
\begin{align*}
\Ex_{t,x,i,z}[U(X^{\pi,l}(T))]\leq \varliminf_{a\to\infty}\Ex_{t,x,i,z}[U(X^{\pi,l}(T\wedge\tau_a))]\leq x^\gamma \varphi(t,i,z).
\end{align*}
This verifies the validity of the conclusion (i).

We next prove the conclusion (ii). In fact, recall that the optimal feedback strategy $(\pi^{*},l^*)=(\pi^*(t,i,z),l^*(t,i,z))$ for $i=1,\ldots,m$ is given by \eqref{eq:lnstar}
for $k=n$ and given by \eqref{eq:optimumk} for $k=0,1,\ldots,n-1$. Then, there exists a constant $C>0$ which is independent of $(t,i,z)$ such that
$\|\pi^*(t,i,z)\|^2+|l^*(t,i,z)|^{2}\leq C$ for all $(t,i,z)\in[0,T]\times\{1,\ldots,m\}\times{\cal S}$. We next estimate $\Ex[(X^{\pi^*,l^*}(T\wedge\tau_a))^{2\gamma}]$.
First of all, the dynamics of the wealth process can be rewritten as, for $s\in[t,T]$,
\begin{align*}
&dX^{\pi^*,l^*}(s)=X^{\pi^*,l^*}(s)[r(Y_{s})+\pi^{*}(s,Y(s),Z(s))^{\top}\theta(Y(s),Z(s))\nonumber\\
&\quad+l^*(s,Y(s),Z(s))(p(Y(s),Z(s))-c(Y(s)))]ds\nonumber\\
&\quad+X^{\pi^*,l^*}(s)[\pi^{*}(s,Y(s),Z(s))^{\top}\sigma(Y(s))-l^*(s,Y(s),Z(s))\phi(Y(s))]dW(s)\nonumber\\
&\quad-X^{\pi^*,l^*}(s)l^*(s,Y(s),Z(s))\bar{\phi}(Y(s))d\bar{W}(s)\nonumber\\
&\quad-X^{\pi^*,l^*}(s-)\pi^{*}(s,Y(s),Z(s))^{\top}dZ(s)-l^*(s-,Y(s-),Z(s-))X^{\pi^*,l^*}(s-)g(Y(s-))dN(s).
\end{align*}
Then, It\^o's formula yields that for $u\in[t,T]$,
\begin{align*}
(X^{\pi^*,l^*}(u))^{2\gamma}=&(X^{\pi^*,l^*}(t))^{2\gamma}+\tilde{M}^{\pi^*,l^*}(u)-\tilde{M}^{\pi^*,l^*}(t)\nonumber\\
&+\int_t^u(X^{\pi^*,l^*}(s))^{2\gamma}\tilde{\cal A}(\pi^*(s,Y(s),Z(s)),l^*(s,Y(s),Z(s));Y(s),Z(s))ds.
\end{align*}
Here, for $(\pi,l)\in(-\infty,1]^n\times[0,\infty)$ and $(i,z)\in\{1,\ldots,m\}\times{\cal S}$,
\begin{align*}
\tilde{\cal A}(\pi,l;i,z)=&{2\gamma}\big[r(i)+\pi^{\top}(I-diag(z))\theta(i,z)+(p(i,z)-c(i))l\big]\\
&+{\gamma}({2\gamma}-1)\big[\pi^{\top}(I-diag(z))\sigma(i)\sigma(i)^{\top}(I-diag(z))\pi+l^2\big(\phi(i)\phi(i)^{\top}+\bar{\phi}(i)\bar{\phi}(i)^{\top}\big)\\
&-2l\pi^{\top}(I-diag(z))\sigma(i)\phi(i)^{\top}\big]+[(1-lg(i))^{2\gamma}-1]\nu(i,z)\\
&+\sum_{j=1}^n[(1-\pi_j)^{2\gamma}-1](1-z_j)h_j(i,z).
\end{align*}
The $\Px$-(local) martingale is given by, for $t\in[0,T]$,
\begin{align*}
\tilde{M}^{\pi^*,l^*}(t):=&\int_0^t(X^{\pi^*,l^*}(s))^{2\gamma}[\pi^{*}(s,Y(s),Z(s))^{\top}\sigma(Y(s))-l^*(s,Y(s),Z(s))\phi(Y(s))]dW(s)\nonumber\\
&-\int_0^t(X^{\pi^*,l^*}(s))^{2\gamma}l^*(s,Y(s),Z(s))\bar{\phi}(Y(s))d\bar{W}(s)\nonumber\\
&+\sum_{j=1}^n\int_0^t(X^{\pi^*,l^*}(s-))^{2\gamma}[(1-\pi_j^*(s,Y(s-),Z(s-)))^{2\gamma}-1]dM_j(s)\nonumber\\
&+\int_0^t(X^{\pi^*,l^*}(s-))^{2\gamma}[(1-lg(Y(s-)))^{2\gamma}-1]d\tilde{N}(s).
\end{align*}
As above, we have that $\|\pi^*(t,i,z)\|^2+|l^*(t,i,z)|^{2}\leq C$ for all $(t,i,z)\in[0,T]\times\{1,\ldots,m\}\times{\cal S}$, and hence
\[
|(1-l^*(t,i,z)g(i))^{2\gamma}-1|\leq(1+\gamma)(|l^*(t,i,z)g(i)|^2+|l^*(t,i,z)g(i)|).
\]
Then, there exists a constant $C>0$ such that for all $(t,i,z)\in[0,T]\times\{1,\ldots,m\}\times{\cal S}$,
\[
|\tilde{\cal A}(\pi^*(t,i,z),l^*(t,i,z),i,z)|\leq C.
\]
Thus, we have that for all $t\in[0,T]$,
\begin{align*}
&\Ex_{t,x,i,z}\big[(X^{\pi^*,l^*}(T\wedge\tau_a))^{2\gamma}\big]
=x^{2\gamma}\nonumber\\
&\qquad+\Ex_{t,x,i,z}\left[\int_t^{T\wedge\tau_a}(X^{\pi^*,l^*}(s))^{2\gamma}\tilde{\cal A}(\pi^*(s,Y(s),Z(s)),l^*(s,Y(s),Z(s));Y(s),Z(s))ds\right]\\
&\quad\leq x^{2\gamma}+\Ex_{t,x,i,z}\left[\int_t^{T}(X^{\pi^*,l^*}(s\wedge\tau_a))^{2\gamma}\big|\tilde{\cal A}(\pi^*(s,Y(s),Z(s)),l^*(s,Y(s);Z(s)),Y(s),Z(s))\big|ds\right]\\
&\quad\leq x^{2\gamma}+C\int_t^{T}\Ex_{t,x,i,z}[(X^{\pi^*,l^*}(s\wedge\tau_a))^{2\gamma}]ds.
\end{align*}
The Gronwall's inequality yields that
\[
\sup_{a\in\R_+}\Ex_{t,x,i,z}\big[(X^{\pi^*,l^*}(T\wedge\tau_a))^{2\gamma}\big]\leq x^{2\gamma}e^{CT},
\]
and hence $\{(X^{\pi^*,l^*}(T\wedge\tau_a))^{\gamma}\}_{a\in\R_+}$ is uniformly integrable. This yields that
\begin{align*}
V(t,x,i,z)=\Ex_{t,x,i,z}[U(X^{\pi^*,l^*}(T))]=\lim_{a\to\infty}\Ex_{t,x,i,z}[U(X^{\pi^*,l^*}(T\wedge\tau_a))]= x^\gamma \varphi(t,i,z).
\end{align*}
This verifies the validity of the conclusion (ii). \hfill$\Box$

\end{document}